\documentclass[fleqn,usenatbib,useAMS]{mnras}

\usepackage{graphicx}	
\usepackage{amsmath}	
\usepackage{amssymb}	
\usepackage{multicol}        
\usepackage{bm}		
\usepackage{pdflscape}	
\usepackage{subfigure}

\newcommand{\km}{\rm\thinspace km}

\newcommand{\s}{\rm\thinspace s}
\newcommand{\ks}{\rm\thinspace ks}

\newcommand{\Ms}{\rm\thinspace Ms}

\newcommand{\Hz}{\rm\thinspace Hz}

\newcommand{\Msun}{\hbox{$\rm\thinspace M_{\odot}$}}

\newcommand{\keV}{\rm\thinspace keV}

\newcommand{\kmps}{\hbox{$\km\s^{-1}\,$}}

\newcommand{\rg}{\rm\thinspace $r_\mathrm{g}$}

\usepackage[T1]{fontenc}
\usepackage{ae,aecompl}

\usepackage{newtxtext,newtxmath}
\title[Wavelet spectral timing]{Wavelet spectral timing: X-ray reverberation from a dynamic black hole corona hidden beneath ultrafast outflows}

\author[D. R. Wilkins]{D. R. Wilkins\thanks{Contact e-mail: \href{mailto:dan.wilkins@stanford.edi}{dan.wilkins@stanford.edu}}
\\
\hspace{-0.25em}Kavli Institute for Particle Astrophysics and Cosmology, Stanford University, 452 Lomita Mall, Stanford, CA 94305, USA
}

\date{Accepted 2023 September 22. Received 2023 September 11; in original form 2023 June 25}

\pubyear{2023}

\begin{document}
\label{firstpage}
\pagerange{\pageref{firstpage}--\pageref{lastpage}}
\maketitle

\begin{abstract}
Spectral timing analyses based upon wavelet transforms provide a new means to study the variability of the X-ray emission from accreting systems, including AGN, stellar mass black holes and neutron stars, and can be used to trace the time variability of X-ray reverberation from the inner accretion disc. The previously-missing iron K reverberation time lags in the AGN IRAS\,13224$-$3809 and MCG--6-30-15 are detected and found to be transitory in nature. Reverberation can be hidden during periods in which variability in the iron K band becomes dominated by  ultrafast outflows (UFO). Following the time evolution of the reverberation lag between the corona and inner accretion disc, we may observe the short-timescale increase in scale height of the corona as it is accelerated away from the accretion disc during bright X-ray flares in the AGN I\,Zw\,1. Measuring the variation of the reverberation lag that corresponds to the continuous, stochastic variations of the X-ray luminosity sheds new light on the disc-corona connection around accreting black holes. Hysteresis is observed between the X-ray count rate and the scale height of the corona, and a time lag of 10$\sim$40\ks\ is observed between the rise in luminosity and the increase in reverberation lag. This correlation and lag are consistent with  viscous propagation through the inner accretion disc, leading first to an increase in the flux of seed photons that are Comptonised by the corona, before mass accretion rate fluctuations reach the inner disc and are able to modulate the structure of the corona.
\end{abstract}




\section{Introduction}
In recent years, great advances have been made in our understanding of the extreme environments just outside the event horizons of accreting black holes with the advent of spectral-timing analysis, combining measurements of the X-ray spectrum, its variability, and the causal connections between different components that contribute to the observed emission. Most notable is the detection of X-ray reverberation from the inner regions of the accretion disc \citep{fabian+09, reverb_review, cackett_reverb_review}. As material accretes onto a black hole (whether that be a stellar mass black hole in an X-ray binary, or a supermassive black hole in an active galactic nucleus, or AGN), a corona of accelerated particles is formed in the inner regions of the accretion flow. This corona goes on to produce the luminous X-ray continuum that we observe, likely by the Compton scattering of lower energy seed photons emitted thermally from the accretion disc \citep{galeev+79,haardt+91}.

A fraction of the continuum photons emitted from the corona illuminate the inner regions of the accretion disc, where they are reprocessed through a combination of Compton scattering, photoelectric absorption, fluorescent line emission and bremsstrahlung radiation, giving rise to a characteristic `reflection' spectrum. This reflection spectrum contains a number of emission lines, including the prominent iron K$\alpha$ fluorescent line \citep{george_fabian}, emitted in the rest frame of the material between 6.4\keV\ (in the case of neutral iron) and 6.97\keV\ (for highly-ionised iron). The reflection spectrum and emission lines we observe from the disc are subject to Doppler shifts, from the orbital motion of the accretion disc, and gravitational redshifts, as the photons climb out of the gravitational potential of the black hole to reach us. This results in the emission lines from the accretion disc being broadened into a characteristic shape, with a blueshifted peak and a redshifted wing extending to low energy, consisting of the photons from the innermost regions of the accretion disc \citep{fabian+89}. Below 1\,keV, this relativistic broadening causes a number of emission lines to be blended into a \textit{soft excess} of emission above the power law continuum.

The X-ray emission from the corona is highly variable, and the reflection from the accretion disc responds to changes in the intensity of the primary continuum. Due to the additional light travel time between the corona and disc, variations in the reflection lag behind correlated variations in the continuum. To date, these \textit{reverberation time lags} have been measured around a sample of approximately 20 supermassive black holes in AGN, predominantly in nearby Seyfert galaxies \citep{demarco+2012,kara_global}, in low mass AGN \citep{mallick+2021}, and around a growing number of stellar mass black holes \citep{demarco+2015,kara+2019,wang_reverb_machine}.

The measured time lags scale with the mass of the black hole and correspond to the light crossing time over just $2\sim 10$\rg\ in the case of the supermassive black holes, where the gravitational radius, $r_\mathrm{g} = GM/c^2$ is the characteristic scale length in the gravitational field, and corresponds to the radial co-ordinate of the event horizon  of a maximally spinning black hole. Combining measurements of reverberation time lags with the relativistic energy shifts of emission line photons allows us to map the environment around the black hole and the innermost regions of the accretion flow, enabling measurements of the spin of the black hole \citep{brenneman_reynolds,reynolds_spin_review}, as well as the location, geometry and structure of the X-ray emitting corona \citep{understanding_emis_paper, propagating_lag_paper, 1zw1_corona_paper}, and the structure of the accretion disc itself \citep{taylor_reynolds}.

X-ray spectral timing analysis and the measurement of X-ray reverberation time lags is predominantly based upon Fourier transforms of light curves describing the variability of the X-rays emitted in energy bands that are dominated by the continuum and by reflection from the accretion disc, \textit{i.e.} the soft excess and the iron K line \citep{reverb_review}, or in some cases by fitting Fourier-derived models to the observed data points in unevenly sampled light curves with gaps \citep{zoghbi_gaps}. A Fourier transform decomposes an observed signal into a sum of sinusoidal components, which, by their nature, are global functions that are described by just an amplitude and phase, which do not vary with time (or generalised location within the signal). This means that a Fourier analysis implicitly assumes stationarity of the signal in time and is unable to capture time-variability of the process we are observing. This is not to say that the X-ray emission is constant (we are measuring the variability of the emission), rather it is the properties of that variability that are assumed to be constant during a Fourier analysis, including the power spectrum and the causal relationships and time delays between components.

We know, however, that the X-ray emitting coron\ae\ around black holes are dynamic systems that evolve in time. Long observing campaigns reveal how the power spectrum of the variability of the coronal X-ray emission varies \citep{alston_iras1}, and how the location and geometry of the corona evolves between low and high luminosity states \citep{1h0707_var_paper, mrk335_corona_paper, alston_iras2, caballero-garcia+2020}. In addition to the continuous, stochastic variability of the X-ray luminosity and the corona, transient events such as X-ray flares are observed from around supermassive black holes in AGN \citep[\textit{e.g.}][]{1zw1_nature} and evidence is emerging that the corona cools and is accelerated away from the black hole and accretion disc during bright flares \citep{mrk335_flare_paper, 1zw1_flare_paper}. It is therefore apparent that in order to fully characterise the variability of the corona and to yield the maximum information about the inner accretion flow from X-ray reverberation, we require an analysis technique that is able to capture the non-stationarity of the underlying process.

We might therefore explore \textit{wavelet} analysis as an alternative to Fourier analysis as the basis for spectral timing. Wavelet transforms decompose the observed signal into basis functions that are localised in time (or space) and can therefore be used to follow the evolution of the system in time. Wavelet analysis has found a broad array of applications across fields from the analysis of climate and meteorological data \citep{lau+1995}, to many branches of biomedical science, communications, noise reduction and speech recognition \citep{wavelet_speech}. Within X-ray astronomy, wavelet analysis has found applications detecting features in images, most notably underlying the \textsc{wavdetect} source detection algorithm commonly applied to \textit{Chandra} observations \citep{wavdetect}.  \citet{lachowicz+2005} applied wavelet analysis to X-ray timing, tracing the time-variability of the power spectrum of the black hole X-ray binary Cygnus X-1, and \citet{ghosh+2023} apply wavelet analysis to measure the variability of the power spectrum of the X-ray emission of a number of AGN. Here we apply wavelet analysis more broadly to X-ray spectral timing and to the measurement of X-ray reverberation.

In \S\ref{sec:wavelet}, the principles of wavelet analysis are outlined, before wavelet spectral timing analysis is applied to the analysis of X-ray reverberation from the accretion discs around supermassive black holes in AGN. \S\ref{sec:data} describes the data used for these investigations. In \S\ref{sec:transitory_fek}, wavelet spectral timing analysis is used to uncover previously-undiscovered X-ray reverberation signals that are masked by a variable ultrafast outflow, and in \S\ref{sec:evolution}, wavelet spectral timing analysis is used to follow the evolution of the corona in time during X-ray flares and to explore the disc-corona connection.

\section{From Fourier analysis to wavelet analysis}
\label{sec:wavelet}
A Fourier transform decomposes an observed signal, $s(t)$, into a sum of sinusoidal components, $\psi_\omega = a_\omega\cos(\omega t + \varphi_\omega)$. In the case of timing analyses, or spectral-timing analyses, that signal is a light curve, \textit{i.e.} a time series representing the observed flux or count rate recorded in some energy band as a function of time (discretised in a series of time bins). Each of the sinusoidal Fourier components represents the components describing the variability on different timescales (corresponding to the frequency, $\omega$\footnote{We define $\omega$ as the angular frequency, related to the linear frequency, $\nu$ by $\omega = 2\pi\nu$.}, of those components), with the low frequency components describing the slow variation in the time series, and the high frequency components describing the rapid components of the variability. Each of the Fourier components has a corresponding amplitude, $a_\omega$, which denotes how much of that component is present in the signal, and a phase, $\phi_\omega$, which shifts each sinusoidal component in time to align with the peaks and troughs on a given timescale in the time series.

Each of the Fourier modes is a \textit{global} function, and while each sinusoid represents a component of the time variation of the signal, each of the modes (localised in frequency) exists at all times and has just a single amplitude and phase to reproduce the observed variability. While the Fourier transform provides an exact representation of any periodic signal (or any signal that is bounded at infinity), if one wanted to obtain a physical interpretation of the amplitude and phase of each component (for example to describe some physical process to which the observed emission and its variability can be attributed to), there is an implicit assumption of \textit{stationarity}. That is to say that it is assumed that the underlying process, the values of the amplitudes and phases of individual components, and the phase relationships between the components, remain constant in time. If the underlying variability is non-stationary, the integral that defines the Fourier transform will implicitly average over those quantities.

By contrast, the \textit{wavelet transform} decomposes the observed signal into a sum of components that are localised in both frequency and time, known as \textit{wavelets}. A wavelet can be thought of as a wave packet. Conceptually, the simplest wavelet is the \textit{Morlet} wavelet, which is essentially a sinusoidal carrier wave multiplied by a Gaussian window: 
\begin{align}
	\psi(t) = a_\sigma \pi^\frac{1}{4} e^{-\frac{1}{2}t^2} e^{i\sigma t}
\end{align}
The parameter $\sigma$ represents the width of the wavelet and also encodes the central frequency that it represents: $\omega_\psi =  \sigma / (1 - e^{-\sigma\omega_\psi})$.

The parametrisation of both the width and central frequency of the wavelet using the single $\sigma$ parameter means that one can think of the different frequency components being produced by simply stretching the same basis wavelet (often referred to as the `mother wavelet'). This means that when we consider the wavelet in Fourier space (as a localised wave packet, a single wavelet will be composed of a spread of Fourier components, not just the single Fourier component at the frequency of the underlying carrier wave), the Morlet wavelet has equal variance in both the time and frequency domains. In practice the Morlet wavelet is extremely versatile and can be used to describe a wide variety of signals. The symmetry in both the frequency and time domains, however, means that it cannot always provide the optimal description of the variability in every input signal, particularly for rapid transient events, for which more general wavelet families can be used instead, such as the \textit{Morse} wavelet \citep{olhede+2002}.


\subsection{The wavelet transform}
Just as a Fourier transform decomposes a signal $s(t)$ into an integral over sinusoidal Fourier components representing different frequencies (the Fourier transform at each frequency is a complex number, $\tilde{s}(\omega) = a_\omega e^{i\varphi_\omega}$, representing the amplitude $a_\omega \equiv |\tilde{s}|$ and phase $\varphi_\omega$ of each component), the \textit{wavelet transform} decomposes a signal into an integral over wavelets representing different frequencies ant different points in time. The wavelet transform (specifically the \textit{continuous wavelet transform}), employing a family of wavelets derived from some function $\psi$ (for example the Morlet wavelet, described above), is defined as
\begin{align}
	[W_\psi s](a, b) = \frac{1}{\sqrt{|a|}} \int_{-\infty}^{\infty} \psi\left(\frac{t - b}{a}\right) s(t)\,dt
\end{align}
Notice that the wavelet is not expressed as a function of frequency and time, rather is a function of just $\left(\frac{t-b}{a}\right)$. This definition reflects the fact that the wavelets representing all frequency components are derived by simply stretching a single wavelet function (the `mother wavelet') in time. This rule, that the only allowed transformation of the wavelet is a stretch, is derived from the uncertainty principle in information theory, namely that the uncertainty on frequency and time information are related by $\Delta t \Delta\omega \ge \frac{1}{2}$. Formally, $a$ represents the \textit{scale} of the wavelet, which can also be referred to as its peak or characteristic frequency, $f$, and $b$ represents the variation of each wavelet within the input signal as a function of time. In a continuous wavelet transform, the scale parameter, $a$, can take on arbitrary values, though is typically discretised on a logarithmic scale spaced by a fractional power of 2.\footnote{The continuous wavelet transform is in contrast to the discrete wavelet transform, where the scale parameter, or wavelet frequencies, are discrtetised into strictly integer powers of 2. The discrete wavelet transform ensures that a unique representation of the signal is obtained, and is useful for creating a compressed representation. The continuous wavelet transform, used here, however, is a highly redundant representation of a signal, though is more useful for obtaining physically-interpretable descriptions of the frequency components making up that signal.}

A Fourier transform provides a unique, complete representation of the input function, \textit{i.e.} given the Fourier transform it is possible to exactly reconstruct the time domain signal. The same is true of the wavelet transform, but only if a complete, orthonormal set of wavelets is used. The simplest wavelet forms, the Morlet and the Morse wavelet, do not form orthonormal bases. There is significant overlap in frequency space between wavelets of different scales, thus there is redundancy in the frequency components. That being said, in spectral-timing analysis, we seldom wish to reconstruct the input signal. Rather we seek to use the Fourier transform, or in this case the wavelet transform, to measure the different time scale or frequency components that make up the variability we observe and to use these components to gain an understanding of the underlying physical processes. In the latter case, the Morlet (or Morse) wavelet is preferred over orthonormal wavelets in order to obtain a more intuitive description of the frequency components within the signal.

Just as for a Fourier transform, the wavelet transform of a given signal is valid only over a limited frequency range. For an input signal in discrete time bins of width $\Delta t$, the upper bound on frequency is defined by the same Nyquist limit applicable to a Fourier transform ($\nu_\mathrm{max} = 1/2\Delta t$). The lowest frequency component is limited by the length, $T$, of the input signal, with the requirement that at least one period of the wave must fit within the observing time ($\nu_\mathrm{min} = 1/T$). For a wavelet transform, the latter criterion becomes time-dependent. The wavelet transform at a given time is defined by wave packets centred upon that time. This means that half of one wave cycle must fit between the time bin of interest and the edge of the observation, defining the \textit{cone of influence}, where lower frequency wavelets become invalid towards the edges of the observed time window.

\subsection{The scalogram and the power spectrum}
As for the Fourier transform, the wavelet transform at each scale (which can be translated into frequency) and at each point in time, is a complex number, representing both the amplitude and the phase of each frequency component at that time. The square of the amplitude of the wavelet transform is referred to as the \textit{scalogram} and is analogous to the periodogram in a Fourier analysis. The scalogram represents the power within each frequency component as a function of time and can be used to trace the variation of the power spectrum or the time evolution of spectral features \citep{lachowicz+2005}. Applications of the wavelet scalogram in X-ray timing analysis include tracing the time variability of a quasi-periodic oscillation or QPO \citep{czerny+2010, ghosh+2023}, and tracing the variation of the power spectrum during quasi-periodic erruption from an AGN \citep{ghosh+2023}.

The scalogram is comparable to the Fourier dynamical power spectrum (sometimes referred to in signal processing as the spectrogram, derived from a windowed Fourier transform, or WFT), also known as a short-time Fourier transform (STFT), which is calculated by computing the Fourier transform in short segments of the observation. The WFT or STFT can only trace the variability from window to window, and the window size is fixed. On the other hand, the wavelets of different scales effectively tile the time-frequency plane, such that the higher frequency components of the variability are sampled in shorter time windows. Higher frequency components may vary more quickly than the lower frequency components, and as such, the wavelet scalogram has been shown to be more sensitive to rapid transient events, as the rapid variation of the high frequency components is more accurately captured by the continuous wavelet transform.\footnote{\url{https://www.mathworks.com/help/wavelet/ug/time-frequency-analysis-and-continuous-wavelet-transform.html}}

\subsection{The wavelet coherence and time lags}
A key quantity in spectral timing analysis is the cross spectrum between two signals. The cross spectrum encodes the causal relationship between two signals, containing both the power spectrum of the part of the variability that is correlated between the two time series (the cross power) and the phase or time delay between correlated variations \citep{reverb_review,zoghbi+09}. 

For two time series, $s_1(t)$ and $s_2(t)$, and their respective Fourier transforms, $\tilde{s}_1(\omega) = |\tilde{s}_1|e^{i\varphi}$ and $\tilde{s}_2(\omega) = |\tilde{s}_2|e^{i\vartheta}$, the cross spectrum is given by
\begin{align}
	[\tilde{C}s_1s_2](\omega) = \tilde{s}^*_1 \tilde{s}_2 = |\tilde{s}_1||\tilde{s}_2|e^{i(\vartheta - \varphi)}
\end{align}
where the amplitudes and phases are functions of frequency. 

The argument of the cross spectrum gives the phase lag between the two signals, which can be converted to a time lag for the frequency component in question. In terms of the linear frequency, the time lag is given by:
\begin{align}
	\tau(\nu) = \frac{\arg \tilde{C}}{2\pi\nu}
\end{align}

The degree to which two signals are correlated is expressed by the \textit{coherence}:
\begin{align}
	\gamma^2(\omega) = \frac{|\langle \tilde{s}^*_1 \tilde{s}_2 \rangle|^2}{\langle|\tilde{s}_1|^2 \rangle \langle|\tilde{s}_2|^2 \rangle}
\end{align}
The coherence is defined in frequency \textit{bins} and in the case of the Fourier coherence, angle brackets denote the average of each quantity across the frequency components within each bin. Formally, the coherence represents the fraction of the variability in one of the time series that can be predicted from a linear transformation of the variability observed in the other, and takes a value between 0 for uncorrelated signals, and 1 for perfectly correlated signals, and can be used to estimate the error on the lag measurement \citep{nowak+99}.

Similarly, the coherence can be defined in a wavelet analysis, referred to as the \textit{wavelet coherence}, $R_\psi$:
\begin{align}
	[R_\psi s_1 s_2](f, t) = \frac{\langle [W_\psi s_1]^*[W_\psi s_2] \rangle}{\sqrt{\langle|[W_\psi s_1]|^2 \rangle \langle|[W_\psi s_2]|^2 \rangle}}
\end{align}
In this case, angle brackets denote a generalised smoothing operator. While a Fourier coherence is computed in frequency bins, in wavelet analyses it is typical to take the moving average across both time and wavelet scale or frequency points as the smoothing operator.\footnote{Note that we use $f$ to denote the characteristic or peak frequency represented by the wavelet, defined by its scale, which in reality as a short timescale wave packet contains multiple Fourier frequency components, by contrast with $\nu$ which denotes the Fourier frequencies.} A moving average filter smoothing across 12 scales and 12 time bins was used in this analysis.

It should be noted that while the Fourier coherence is defined as the squared magnitude, and is a real number between 0 and 1, by convention the wavelet coherence is complex. The numerator of the Fourier coherence is the squared-magnitude of the cross spectrum (averaged over the frequency bin), however the numerator of the wavelet coherence is the wavelet cross spectrum $[C_\psi s_1 s_2](f, t) = [W_\psi s_1]^*[W_\psi s_2]$, with the moving average or smoothing operator applied, containing both the amplitude and phase lag information. The equivalent of the coherence scalar is the squared magnitude of the wavelet coherence, \textit{i.e.} $\gamma_\psi^2 = |[R_\psi s_1 s_2]|^2$.

Using this definition, the phase lags, and hence the time lags, between the two signals, as a function of wavelet frequency and time, can be computed from the wavelet coherence (taking $f$ to be the characteristic or peak frequency represented by each wavelet):
\begin{align}
	\tau(f,t) = \frac{\arg [R_\psi s_1 s_2]}{2\pi f}
\end{align}

Given light curves of the flux or count rate in two energy bands, we can use the wavelet transform and wavelet coherence to calculate the time lags as a function of frequency (or wavelet scale). This is analogous to calculating the lag \textit{vs.} frequency spectrum in a Fourier spectral timing analysis, representing the time delay as one energy band responds to variations in another, separating the low and high frequency components that make up the slow and fast components of the variability. The wavelet time lag spectrum adds a further dimension, time, allowing us to trace how those time lags (and the underlying system) evolve.

In addition to calculating the lag \textit{vs.} frequency spectrum, we can also calculate the time lags between a number of light curves in different energy bands and a common reference band, averaging the lag for each band over a chosen range of frequencies or timescales. This allows us to produce the lag \textit{vs.} energy spectrum, showing the time delays as different energy bands respond to variability from a given process or over given range of timescales/frequencies. We can similarly produce a lag \textit{vs.} energy spectrum using a wavelet analysis, allowing us to trace the variation of the lag-energy spectrum over time.

\subsection{The covariance and the variable part of the spectrum}
The \textit{covariance} spectrum reveals the part of the spectrum\footnote{Here we refer to the flux spectrum, \textit{i.e.} the flux or count rate as a function of photon energy or frequency, as opposed to a spectrum in the timing sense.} that is variable within a given range of frequencies or timescales \citep{reverb_review}. The covariance is calculated from the cross spectrum between the light curve in each energy band and a common reference band, and, as such, picks out only the component of the variability that is correlated between each band and the reference. The reference band therefore acts as a matched filter and reduces the noise from uncorrelated components, unlike a traditional RMS spectrum.

By analogy with the Fourier covariance spectrum, we may define the wavelet covariance between the time series in one energy band, $s_E$, and that in a reference band $s_\mathrm{ref}$, in order to trace the time variability of the variable components within the spectrum \footnote{The factor of $\frac{1}{2}$ gives the wavelet covariance spectrum the same normalisation as the Fourier covariance, in units of the RMS count rate, assuming the conventional L1 normalisation of the wavelets is used.}:
\begin{align}
	[Cv_\psi s_\mathrm{ref} s_\mathrm{E}](f, t) = \frac{1}{2}\sqrt\frac{|[W_\psi s_\mathrm{ref}]^*[W_\psi s_E]|^2}{|[W_\psi s_\mathrm{ref}]|^2}
\end{align}

\subsection{Errors and uncertainties}
To estimate the uncertainty on phase and time lag measurements made via wavelet transforms, Monte Carlo simulations were performed. Each of the observed light curves was resampled 1,000 times by replacing the count rate measured in each bin with a random count rate drawn from a Poisson distribution with mean corresponding to the observed count rate in that bin. The wavelet analyses were then performed on each set of resampled light curves, calculating the lag or coherence spectra for each sample. The distributions of sampled spectra were then used to estimate the confidence limits, and hence the error bars, on each measurement.

We may also follow the geometric formalism adopted in \citet{nowak+99} to derive an analytic estimate of the uncertainty. We consider the complex value of the cross spectrum (either the Fourier cross spectrum or the wavelet cross spectrum/coherence) as a vector in the Argand plane, and the average cross spectrum across some bin in frequency and time as the sum of those vectors.

If the underlying signal is coherent over that bin (\textit{i.e.} the two time series have a consistent phase relationship), the vectors that represent the cross spectrum signal will point in the same direction, with their respective magnitudes corresponding to the amplitude, and their directions corresponding to the phase lag between the two time series. When these vectors are added together to compute the average over the bin, they will add coherently, producing a larger signal vector pointing in that same direction.

The measurement of each of those cross spectral vectors (\textit{i.e.} the measurement at each point in frequency and time) will be accompanied by noise, whether that be the Poisson noise associated with the measurement of the light curves in each time bin, or some other component of each time series that is uncorrelated between the pair. This noise may be represented by an additional vector component that is added to each of the signal vectors. The magnitude of the noise vectors corresponds to the magnitude of the noise in the measurement, however the direction of each noise vector will be random if the noise is uncorrelated with the signal. Adding the sum of the noise vectors to the sum of the signal vectors essentially results in a random walk of the end of the vector around its true location, thus producing the error on the phase of the cross spectrum and, thus, the uncertainty in the phase lag between the two time series.

\citet{bendat_piersol} show that the uncertainty on the phase lag, derived in this way, is given by
\begin{align}
\label{equ:phase_error}
	\Delta\varphi = \sqrt{\frac{1 - \gamma^2}{2\gamma^2 N}}
\end{align}
With the corresponding error on the time lag given by $\Delta\tau = \Delta\varphi / 2\pi f$. Following the above derivation and geometric picture, one can see that this same formula is applicable to both Fourier and wavelet lag measurements, simply using the squared-magnitude of the wavelet coherence for $\gamma^2$. The $N$ in the denominator is equal to the number of frequency points that are averaged over in the bin (often $N=KM$ for $K$ distinct Fourier frequencies from $M$ light curve segments). In the wavelet coherence and lag calculation, when tracing the variability over timescales relatively short compared to the width of the moving average filter along the time axis, $N$ correspond to the width of the moving average filter, or the number of bins over which the wavelet coherence is averaged on the frequency axis ($N = 12$ in this case). If the wavelet coherence or lag is averaged over a large number of time bins (significantly larger than the width of the moving average filter along the time axis), such as for the calculation of the lag-energy spectrum, the phase uncertainty tends towards the value for N corresponding to the product of the width of the moving average filters along the time and frequency axes ($N = 12^2 = 144$).

This analytic approximation provides a good approximation to error derived from the Monte Carlo simulations in the regime that the coherence is high. Simulations show that coherence values of $\gamma^2 \gtrsim 0.25$ are required to accurately estimate the uncertainty of the phase lag using Equation~\ref{equ:phase_error} (see Appendix~\ref{app:errors} for further details). It should be noted that this formula holds true not just for uncertainty induced by Poisson noise, but for any uncorrelated component of between the pair of light curves, which will serve to reduce the measured coherence.

\section{Observations and data reduction}
\label{sec:data}
To explore the application of wavelet spectral timing analyses to X-ray observations of black holes, these techniques were applied to \textit{XMM-Newton} observations of the AGN IRAS\,13224$-$3809, MCG--6-30-15, and I\,Zw\,1, selected due to the long available exposures, and the interesting behaviours that have previously been observed in these AGN, described in the later sections. The observations are detailed in Table~\ref{tab:data}.

\begin{table}
		\caption{The \textit{XMM-Newton} observations of the AGN IRAS\,13224$-$3809, MCG--6-30-15 and I\,Zw\,1 used in this analysis.}
	\begin{center}		
		\begin{tabular}{lllc}
		\hline
		Target & OBSID & Start Date & Exposure \\
		\hline
		IRAS\,13224$-$3809 & 0110890101 & 2002-01-19 & 64\ks \\
		& 0673580101 & 2011-07-19 & 133\ks \\
		& 0673580201 & 2011-07-21 & 132\ks \\
		& 0673580301 & 2011-07-25 & 129\ks \\
		& 0673580401 & 2011-07-29 & 135\ks \\
		& 0780560101 & 2016-07-08 & 141\ks \\
		& 0780561301 & 2016-07-10 & 141\ks \\
		& 0780561401 & 2016-07-12 & 138\ks \\
		& 0780561501 & 2016-07-20 & 141\ks \\
		& 0780561601 & 2016-07-22 & 141\ks \\
		& 0780561701 & 2016-07-24 & 141\ks \\
		& 0792180101 & 2016-07-26 & 141\ks \\ 
		& 0792180201 & 2016-07-30 & 141\ks \\
		& 0792180301 & 2016-08-01 & 141\ks \\
		& 0792180401 & 2016-08-03 & 141\ks \\
		& 0792180501 & 2016-08-07 & 138\ks \\
		& 0792180601 & 2016-08-09 & 138\ks \\
		\hline
		MCG--6-30-15 & 0111570101 & 2000-07-11 & 46\ks \\
		& 0111570201 & 2000-07-11 & 66\ks \\
		& 0029740101 & 2001-07-31 & 89\ks \\
		& 0029740701 & 2001-08-02 & 129\ks \\
		& 0029740801 & 2001-08-04 & 130\ks \\
		& 0693781201 & 2013-01-29 & 134\ks \\
		& 0693781301 & 2013-01-31 & 134\ks \\
		& 0693781401 & 2013-02-02 & 49\ks \\
		\hline
		I\,Zw\,1 & 0851990101 & 2020-01-12 & 76\ks \\
		& 0851990201 & 2020-01-14 & 69\ks \\
		\hline
		\end{tabular}
	\end{center}
	\label{tab:data}
\end{table}

Analysis was conducted primarily on light curves extracted from the EPIC pn camera, due to the instrument's superior sensitivity, particularly when analysing the variability of the X-ray emission, although simultaneous light curves were also extracted from the EPIC MOS cameras. The \textit{XMM-Newton} observations were reduced using the \textsc{xmm science analysis system (sas)} v18.0.0. The event lists were reprocessed using the \textsc{epproc} task, applying the latest available version of the calibration. Source photons were extracted from a circular region, 35\,arcsec in diameter, and the background was extracted from a circular region of the same size, located on the same detector chip. Light curves were constructed from the event lists using \textsc{evselect}, and were corrected to account for dead time and exposure variations using the \textsc{epiclccorr} task.

\section{The transitory iron K reverberation signal}
\label{sec:transitory_fek}
Wavelet spectral timing analysis was used to explore the time variability of X-ray reverberation signals from the inner accretion disc in the AGN IRAS\,13224$-$3809 and MCG--6-30-15. These are both classified as narrow-line Seyfert 1 galaxies that have been noted for strong relativistically broadened iron K lines in their X-ray spectra, attributed to the reflection of the primary X-ray continuum from the inner disc \citep{fabian+2013, tanaka+95}.

Alongside the broad iron K line, reverberation time lags are observed, consistent with the light travel time between a compact corona and the inner disc, are observed. Time delays are observed between the continuum-dominated 1.2-4\keV\ band and the soft, 0.3-1\keV\ X-ray band, dominated by a soft excess in the reflection spectrum, where multiple emission lines (including the iron L line) are blended together by Doppler shifts and gravitational redshifts \citep{fabian+2013,alston_iras2,caballero-garcia+2020,kara+2014}. If the relativistically broadened iron K line and the soft X-ray reverberation lag come from the same reflection from the inner disc, one would expect a similar reverberation time lag to be observed between the continuum and the iron K band \citep{kara_global}. Despite the detection of reverberation in the soft X-ray band, the corresponding iron K reverberation lags are mysteriously not detected in IRAS\,13224$-$3809 and MCG--6-30-15, using conventional Fourier spectral timing techniques that average the cross spectrum and phase lags over the course of long observations. Without the corresponding detection of the iron K reverberation lag, it is difficult to confirm that the soft X-ray lag arises due to reverberation from the inner disc, and not some other variable emission component in the soft X-ray band that is responding to changes in the continuum.

\subsection{Time variability of the reverberation signal}
\begin{figure*}
\centering
\subfigure[IRAS\,13224$-$3809] {
\includegraphics[width=0.49\textwidth]{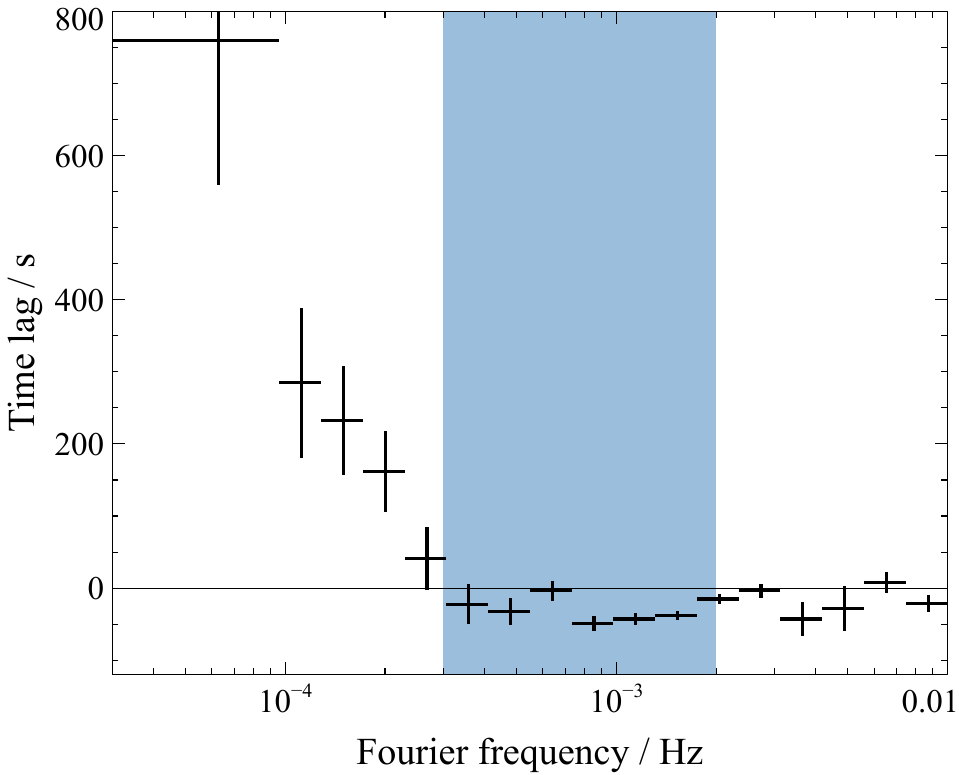}
\label{fig:lagfreq:iras}
}
\subfigure[MCG--6-30-15] {
\includegraphics[width=0.49\textwidth]{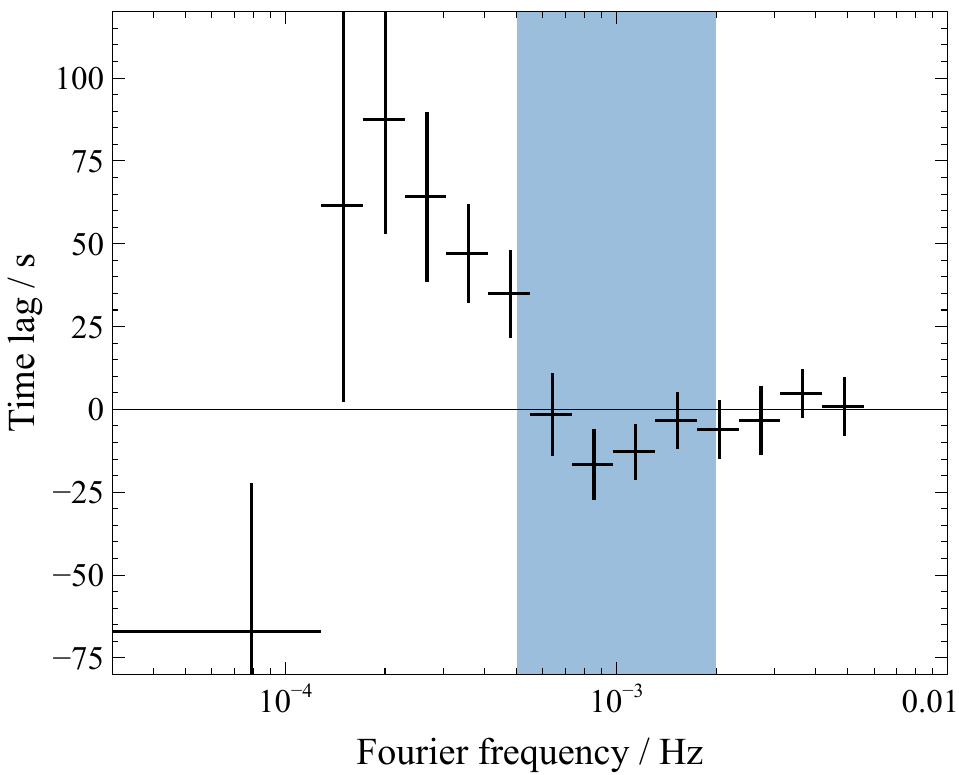}
\label{fig:lagfreq:mcg6}
}
\caption[]{The time lag \textit{vs} Fourier frequency between the 0.3-1\keV\ soft X-ray band and the 1.2-4\keV\ band, dominated by the primary continuum observed in the AGN \subref{fig:lagfreq:iras} IRAS\,13224$-$3809 and \subref{fig:lagfreq:mcg6} MCG--6-30-15. These lag \textit{vs.} frequency spectra were derived using conventional Fourier analysis techniques, and are averaged across the entirety of the observations. X-ray reverberation is detected over frequencies showing a negative time lag, where variations in the 0.3-1\keV\ band, dominated by reflection from the accretion disc, lag behind variations in the continuum.}
\label{fig:lagfreq}
\end{figure*}

The time-averaged lag \textit{vs.} frequency spectra for IRAS\,13224$-$3809 and MCG--6-30-15, calculated using the conventional method, from the Fourier cross spectrum are shown in Fig.~\ref{fig:lagfreq}. From these time-averaged lag spectra, we can identify the ranges of frequencies or timescales upon which reverberation from the disc is observed. Reverberation is where the soft 0.3-1\keV\ band lags behind the 1.2-4\keV\ continuum band, \textit{i.e.} where the soft band lag \textit{vs.} frequency spectrum is \textit{negative} \citep{zoghbi+09}. In IRAS\,13224$-$3809, reverberation is detected over the frequency range $3\times 10^{-4}$ to $2\times 10^{-4}$\Hz\, and in MCG--6-30-15 it is detected over the $5\times 10^{-4}$ to $2\times 10^{-3}$\Hz\
 range. 
 
Below this frequency range, the \textit{hard lag} is detected, where variability in higher energy X-rays systematically lags behind variability in the lower energies. Hard X-ray lags are commonly observed in both AGN and black hole X-ray binaries and are attributed to the propagation of luminosity fluctuations through the corona itself, rather than to reverberation \citep{miyamoto+88,kotov+2001,arevalo+2006}.

\begin{figure*}
\centering
\subfigure[1.2-4\keV\ \textit{vs.} 0.3-1\keV] {
\includegraphics[width=0.49\textwidth,trim=0 0 0 170mm,clip]{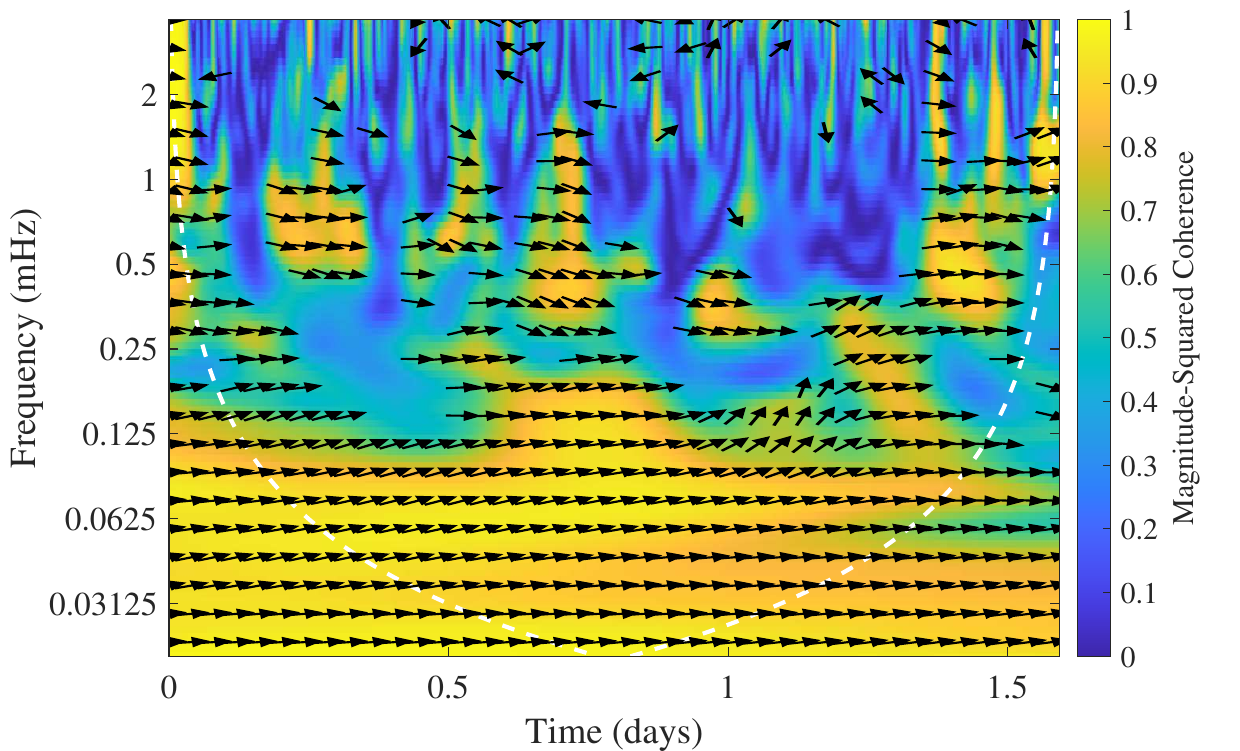}
\label{fig:wavelet_coherence:soft}
}
\subfigure[4-7\keV\ \textit{vs.} 1-4\keV] {
\includegraphics[width=0.49\textwidth,trim=0 0 0 170mm,clip]{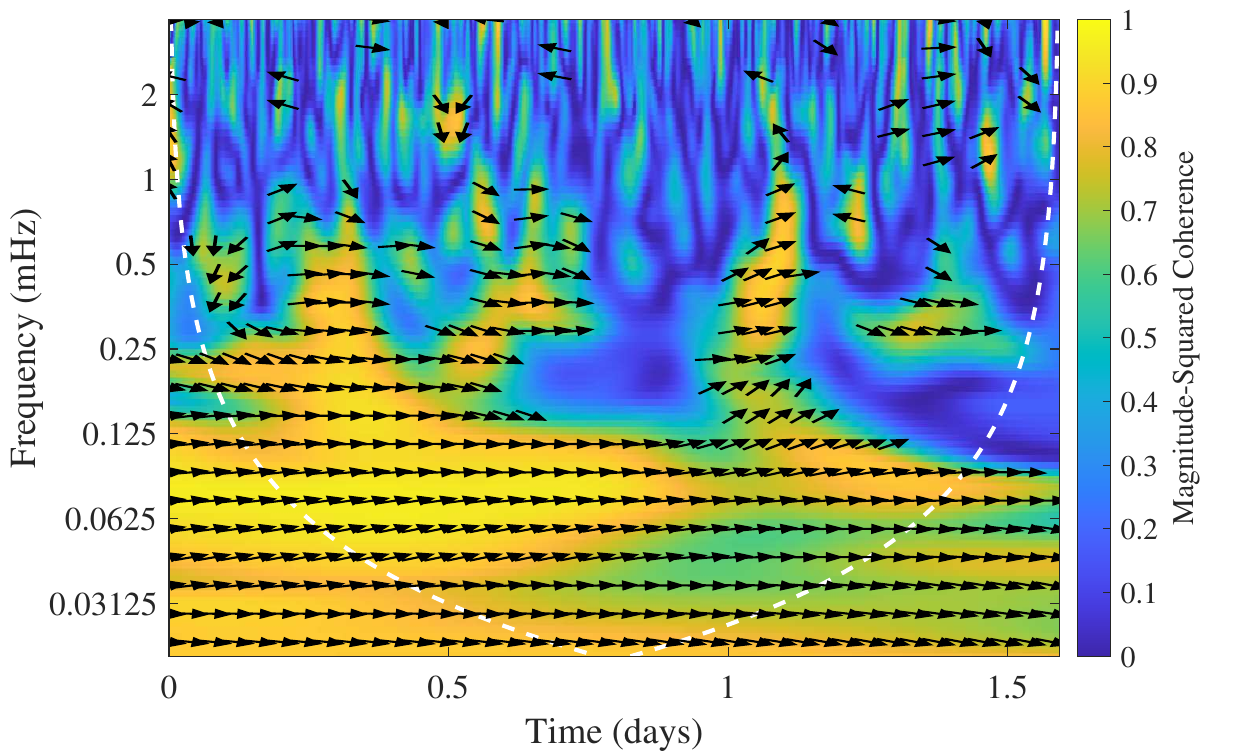}
\label{fig:wavelet_coherence:fek}
}
\caption[]{The wavelet coherence and lag spectra for single observations of IRAS\,13224$-$3809 for \subref{fig:wavelet_coherence:soft} the continuum-dominated 1.2-4\keV\ band \textit{vs.} the 0.3-1\keV\ band dominated by reflection from the disc, and \subref{fig:wavelet_coherence:fek} the 4-7\keV\ iron K band \textit{vs.} the continuum-dominated 1.2-4\keV\ band. Shading indicates the coherence, with values closer to unity indicating a higher level of correlation between variations in the two energy bands. Arrows indicate phase lags between the two light curves. Upward pointing arrows indicate positive lags, where the harder band lags behind the softer band, and downward pointing arrows indicate negative lags. In the soft X-ray band, reverberation from the accretion disc corresponds to negative lags, while in the iron K band, reverberation corresponds to positive lags. The white dashed line shows the cone of influence, within which wavelet values on a given scale are valid and constrained properly by the data.}
\label{fig:wavelet_coherence}
\end{figure*}

Fig.~\ref{fig:wavelet_coherence:soft} shows the wavelet coherence and lag spectrum between light curves in the 0.3-1\keV\ band, dominated by reflection from the accretion disc, and the 1.2-4\keV\ band, dominated by the primary X-ray continuum observed directly from the corona, during single $~125$\ks\ observations made with \textit{XMM-Newton}. The coherence and lag are calculated for a range of wavelet scales (each of which probes a different frequency component of the variability) and as a function of time. The shading corresponds to the coherence with higher values showing that the variability between the two light curves at a given frequency/timescale at a given time and is more closely correlated. Arrows denote the phase lags between the two light curves at each frequency and time, with upward pointing arrows showing that variations in the harder 1.2-4\keV\ band lag behind those in the soft 0.3-1\keV\ band. Downward pointing arrows show when variations in the soft band lagging those in the hard band. Fig.~\ref{fig:wavelet_coherence:fek} show the equivalent wavelet coherence and lag spectrum between the 1.2-4\keV\ continuum band and the 4-7\keV\ band, dominated by the broad iron K line.

In the time-resolved wavelet coherence and lag spectra, we see that over the lowest frequencies, the hard lag is stable in time. The coherence is close to unity at all times, and the phase lag of the hard band relative to the soft band are approximately constant (the arrows on the plot are parallel and pointing upwards, corresponding to a positive lag).

\begin{figure*}
\centering
\includegraphics[width=0.95\textwidth]{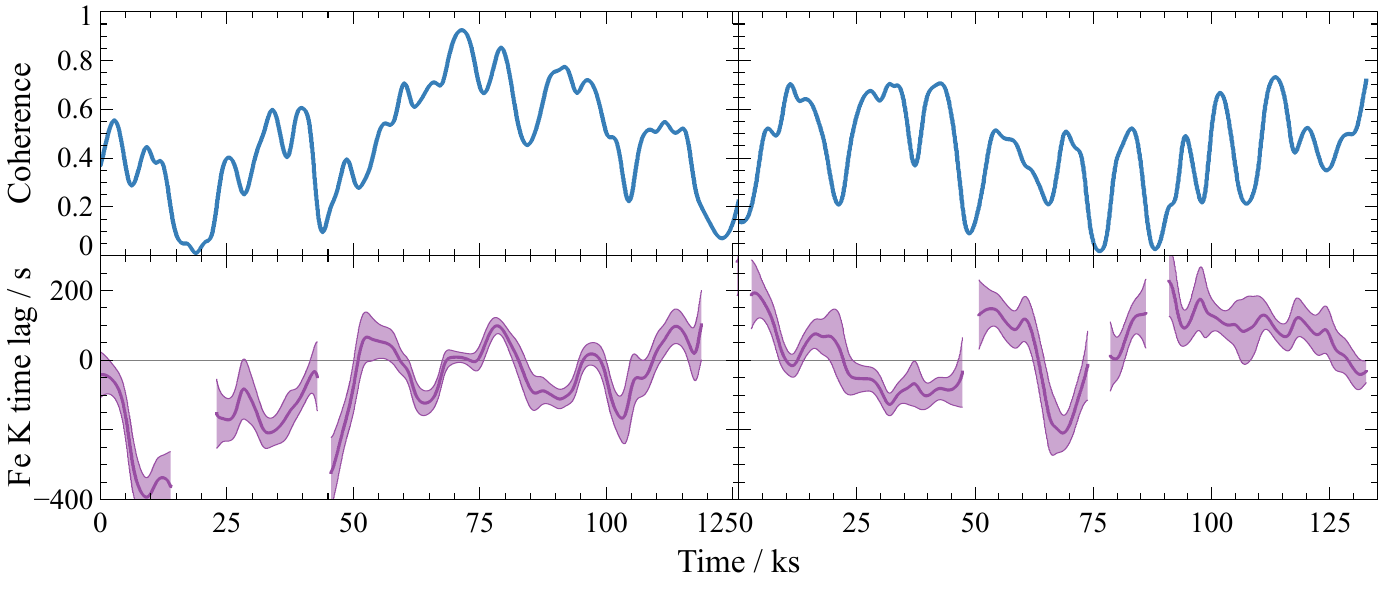}
\caption[]{Time series of the coherence (top panel) and time lag between the variations in the continuum and iron K band (bottom panel) from two of the observations of IRAS\,13224$-$3809. Reverberation from the inner accretion disc is detected during time periods in which the iron K band lags behind the continuum band, indicated by positive values of the time lag. Time lags are undefined when the light curves are incoherent, thus are not shown when the coherence drops below 0.25.}
\label{fig:lag_series}
\end{figure*}

On the other hand, we see that the timing characteristics are much more variable over the high frequency reverberation range. The system transitions through periods of high and low coherence, and the phase lags vary as a function of time. In the soft X-ray band the lags, while variable in their magnitude, remain negative for the majority of the time (\textit{i.e.} the soft X-rays mostly lag behind the continuum-dominated 1.2-4\keV\ band). In the iron K band, however, the lag is much more transitory. Fig.~\ref{fig:lag_series} shows the wavelet coherence and time lag between the iron K and continuum bands, averaged over the $3\times 10^{-4}$-$2\times 10^{-3}$\Hz\ frequency range, as a function of time, over the course of two $\sim 125$\ks\ observations of IRAS\,13224$-$3809. There are time periods in which the iron K emission lags behind the continuum (the lag is positive\footnote{We follow the convention that a positive time lag indicates that the harder X-ray band lags behind the softer X-ray band, thus the reverberation response is defined as positive when the iron K band lags behind the 1-4\keV\ continuum, and negative when the 0.3-1\keV\ soft band lags behind the continuum.}), consistent with reverberation from the inner disc, and also time periods in which emission in the 4-7\keV\ band \textit{leads} the 1.2-4\keV\ continuum band (the lag is negative), as well as time periods in which the two bands show low coherence and a low degree of correlation, summarised in Table~\ref{tab:lag_fraction}.

\begin{table}
		\caption{The fraction of the total observing time during which different causal relationships are detected between the iron K and continuum bands in IRAS\,13224$-$3809 and MCG--6-30-15. X-ray reverberation is detected from the inner accretion disc in the iron K band when the light curves in the continuum and iron K band display a high degree of coherence ($\gamma^2 > 0.25$) and variations in the iron K band lag behind those in the continuum band. Incoherent time periods are defined as those periods in which $\gamma^2 < 0.25$.}
	\begin{center}		
		\begin{tabular}{lllc}
		\hline
		Iron K-continuum relationship & \multicolumn{2}{c}{Fraction of observing time}\\
		 & IRAS\,13224 & MCG--6-30-15 \\
		\hline
		Iron K lags continuum & 0.49 & 0.23 \\
		Iron K leads continuum & 0.32 & 0.34 \\
		Iron K and continuum incoherent & 0.19 & 0.43 \\
		\hline
		\end{tabular}
	\end{center}
	\label{tab:lag_fraction}
\end{table}

\subsection{Time-resolved lag-energy spectra}
From the wavelet coherence and lag spectra, we identify three distinct relationships between emission in the 4-7\keV\ band (dominated by the broad iron K line from the inner accretion disc) and the 1.2-4\keV\ band (dominated by continuum emission observed directly from the corona). There are time periods in which the two bands are highly correlated, and variability in the iron K band lags behind the continuum, consistent with reverberation from the accretion disc. There are time periods in which the two bands remain coherent, but the iron K band appears to lead the continuum band. Finally, there are time periods in which the variability in the iron K and continuum-dominated bands is incoherent and does not show a significant degree of correlation.

We can construct the time lag \textit{vs.} energy spectra, averaged over each of these three (non-contiguous) time intervals. The wavelet coherence/lag spectrum was calculated between light curves in 14 approximately logarithmically-spaced energy bands and a common reference band, taken to be the full 0.3-10\keV\ band covered by \textit{XMM-Newton}, but subtracting the current energy band, so as to avoid correlated noise between the bands \citep[as in][]{zoghbi+2014}. We average the coherence and phase lags over frequencies in the range over which reverberation is seen in the soft band lag-frequency spectra, $3\times 10^{-4}$ to $2\times 10^{-4}$\Hz\ for IRAS\,13224$-$3809 and $5\times 10^{-4}$ to $2\times 10^{-3}$\Hz\ for MCG--6-30-15.

\begin{figure*}
\centering
\subfigure[IRAS\,13224$-$3809] {
\includegraphics[width=0.49\textwidth]{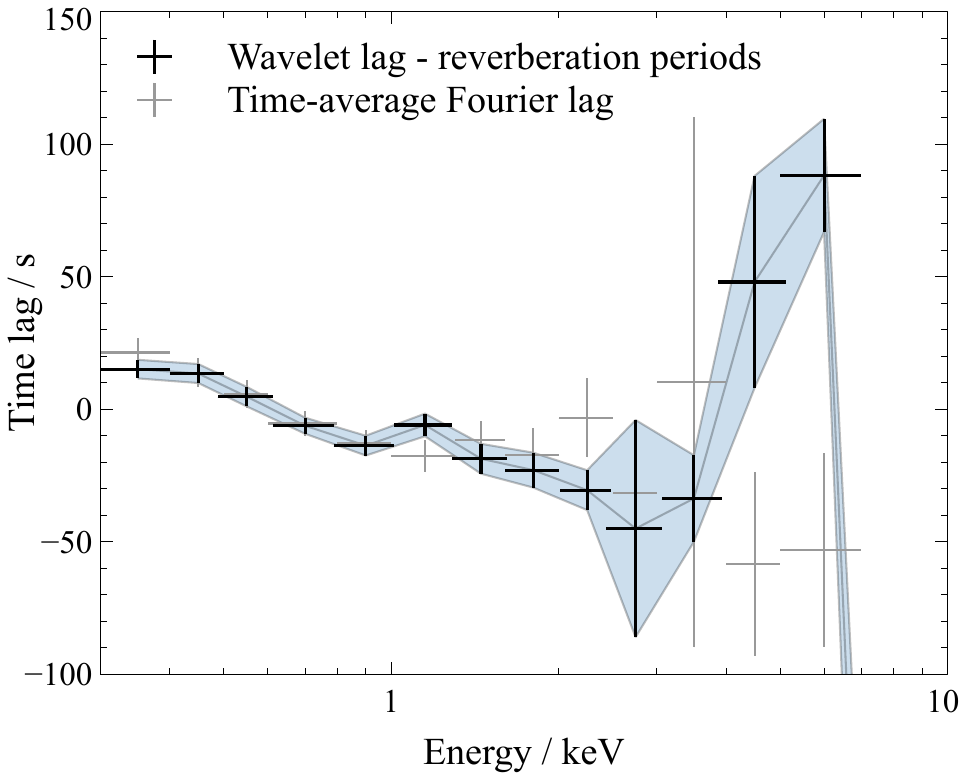}
\label{fig:wavelet_lagen_reverb:iras}
}
\subfigure[MCG--6-30-15] {
\includegraphics[width=0.49\textwidth]{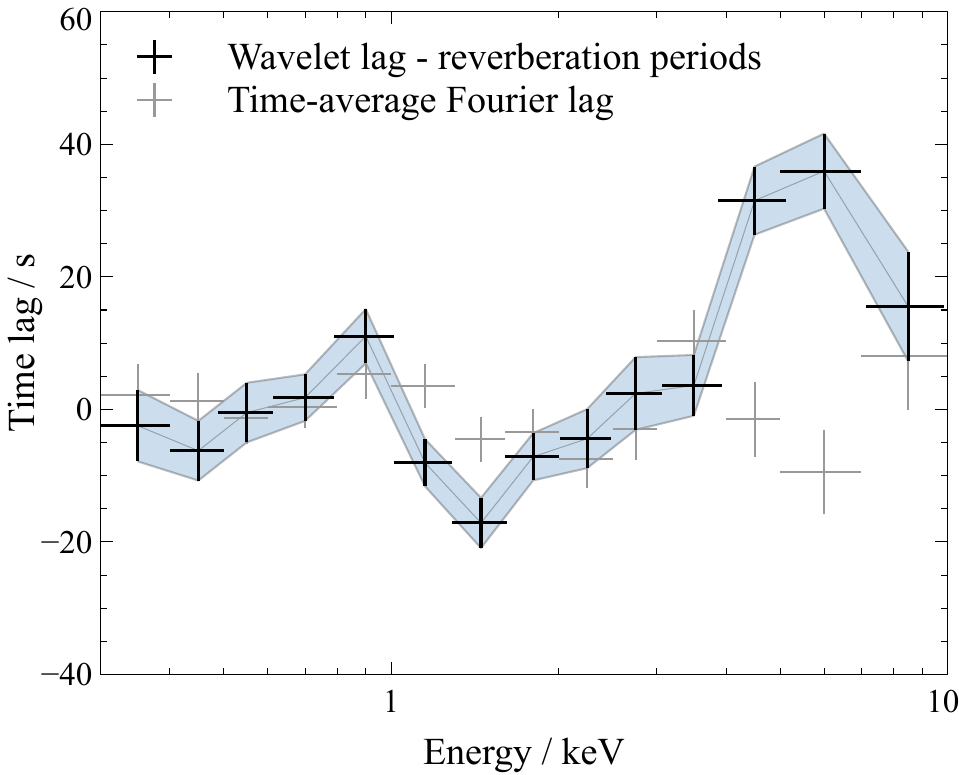}
\label{fig:wavelet_lagen_reverb:mcg6}
}
\caption[]{The lag \textit{vs.} energy spectra, derived from wavelet analysis of \subref{fig:wavelet_lagen_reverb:iras} IRAS$-$13224-3809 and \subref{fig:wavelet_lagen_reverb:mcg6} MCG--6-30-15, selecting only time periods in which the iron K and continuum bands display a high degree of coherence ($\gamma^2 > 0.25$) and in which the variations in the iron K band lag behind those in the continuum band (\textit{i.e.} filtering the time periods during which the reverberation signal is detected). The wavelet-derived lag-energy spectrum is compared to that derived using conventional Fourier analysis techniques, averaged over the entire observations, shown by the light grey data points. The time lags are computed between the light curves in each energy band, and a common reference band, defined to be the full 0.3-10\keV\ band minus the current energy bin, and are averaged over the frequency ranges in which reverberation is detected, highlighted in Fig.~\ref{fig:lagfreq}, and over all time bins in which the coherence is high and the lag is positive.}
\label{fig:wavelet_lagen_reverb}
\end{figure*}

The wavelet cross spectrum was also calculated between the broader 4-7\keV\ iron K band and the 1.2-4\keV\ continuum band, and the coherence and time lag between these two bands, within this same frequency range, as a function of time (Fig.~\ref{fig:lag_series}) was used to filter the energy-resolved wavelet spectra. Time intervals were selected when the iron K and continuum bands were coherent ($\gamma^2 > 0.25$) and variability in the iron K band lags behind that in the continuum (\textit{i.e.} the time lag is positive). The average wavelet coherence (which includes both the scalar coherence and the phase lag) was then calculated over these time periods, and over the chosen ranges of frequencies, to produce the average lag-energy spectrum from this interval, shown in Fig.~\ref{fig:wavelet_lagen_reverb}.

We find that the lag-energy spectrum, averaged over time periods when the iron K band is coherent with, and lags behind, the continuum band, in both IRAS\,13224$-$3809 and MCG--6-30-15, reveals the reverberation signal from the inner accretion disc that is commonly observed among Seyfert-type AGN \citep{kara_global}. Variations in the soft X-ray band, below 1\keV, expected to be dominated by reflection from the accretion disc, as well as in the broad iron K line, between 4 and 7\keV, lag behind correlated variations in the bands most strongly dominated by the continuum (2$\sim$4\keV\ in IRAS\,13224$-$3809 and 1$\sim$2\keV\ in MCG--6-30-15). Variations in the redshifted wing of the iron K line, between $3\sim 5$\keV\ respond sooner than those in the 6\keV\ core of the iron K line, since the most redshifted line emission comes from the innermost radii on the accretion disc, closer to the corona providing the illuminating X-ray continuum (\citealt{zoghbi+2014}, although see \citealt{propagating_lag_paper} for a discussion of the relative timing between the redshifted iron K line and the continuum-dominaed bands). Although this reverberation signal is not detected by Fourier spectral timing methods, averaging over all of the observations, we find that wavelet spectral timing methods are able to recover a time-varying reverberation signal by selecting the time intervals in which the signal is present.


\subsection{Why does the iron K reverberation disappear?}
To understand why the X-ray reverberation signal from the inner accretion disc is detectable only some of the time, we can compare the X-ray flux spectra from the time intervals in which reverberation is detected and time intervals in which it is not.

\begin{figure*}
\centering
\subfigure[IRAS\,13224$-$3809] {
\includegraphics[width=0.49\textwidth]{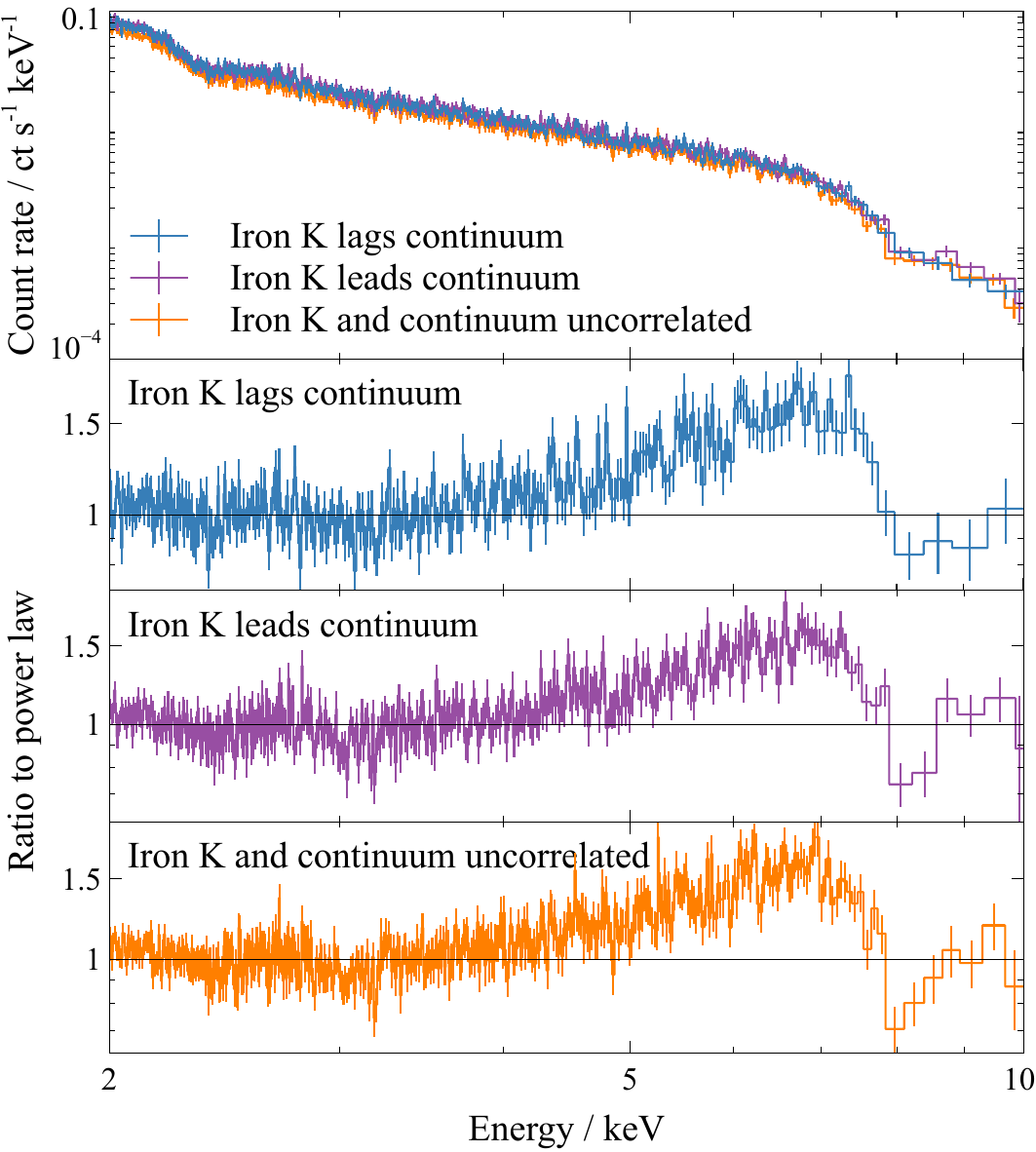}
\label{fig:spectra:iras}
}
\subfigure[MCG--6-30-15] {
\includegraphics[width=0.49\textwidth]{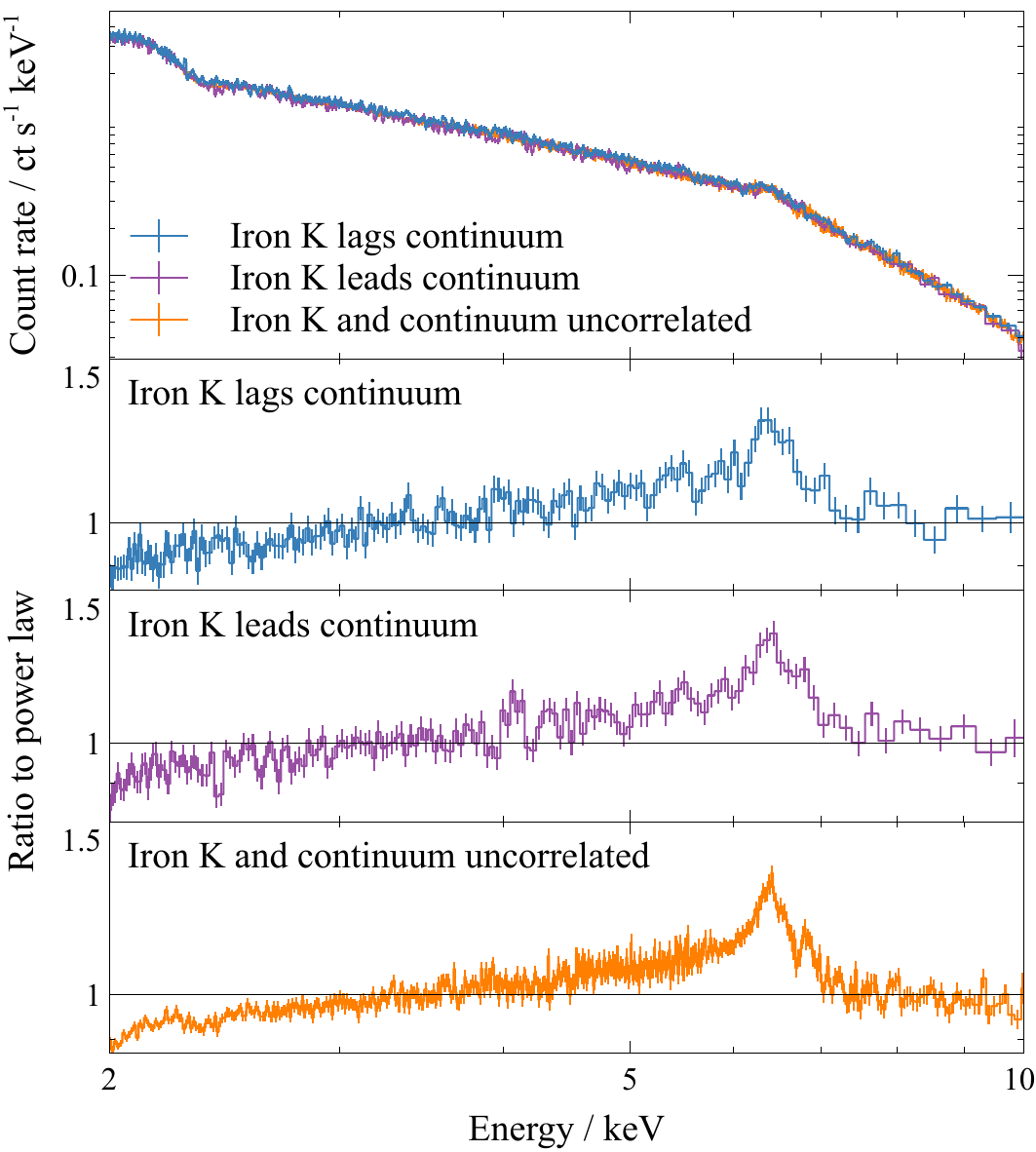}
\label{fig:spectra:mcg}
}
\caption[]{The X-ray flux spectra for \subref{fig:spectra:iras} IRAS\,13224$-$3809 and \subref{fig:spectra:mcg} MCG--6-30-15, extracted using good time interval filters, constructed from wavelet time lag measurements between the iron K and continuum bands. The three spectra for each AGN are averaged across all time intervals in which (1) the iron K band lags behind the continuum and reverberation is detected in the iron K band, (2) the iron K band leads the continuum, and (3) the coherence between the two bands is low. The upper panels show the spectra, while the lower panels show the ratio of the spectra to the best-fitting power law, revealing the shape of the broad iron K line and the absorption from the ultrafast outflows (UFOs).}
\label{fig:spectra}
\end{figure*}

The time series of the coherence and the lag between the iron K and continuum bands were used to construct `good time interval' (GTI) filters, which are used to create spectra from the X-ray photons that were detected during those time intervals. Fig.~\ref{fig:spectra} shows the spectra over the 2-10\keV\ band as well as the ratio of those spectra to the best-fitting power law (which describes the primary continuum spectrum) from the three time intervals: when the iron K band is coherent with and lags behind the continuum (and reverberation is observed), when the iron K band is coherence and leads the continuum band, and when the two bands are incoherent.

In IRAS\,13224$-$3809, we clearly see the profile of the relativistically broadened iron K line in the spectrum, which is most apparent in the ratio between the spectrum and underlying power law. The iron K line has a redshited wing extending to 3$\sim$4\keV, consistent with the reflection from an accretion disc extending to the innermost stable orbit around a rapidly spinning black hole \citep{jiang_iras}.

We observe variation in the absorption feature around 8\keV\ between the time periods in which reverberation is and is not detected. \citet{parker_iras_nature} attribute this absorption feature to an ultrafast outflow (UFO); a wind launched at a velocity of $0.24c$ from the inner accretion disc. The absorption feature is identified as the Fe\,\textsc{xxvi} Ly\,$\alpha$ transition (at 6.97\keV\ in the rest frame), blueshifted as the absorbing material outflows along our line of sight. This absorption feature is weakest during the time periods when reverberation can be seen from the inner disc, and strongest in the time periods when either the iron K band leads the continuum band, or the two bands are incoherent.

In MCG--6-30-15, there is not such a pronounced absorption feature from an ultrafast outflow, and the difference between the flux spectra between the time periods with and without reverberation is less clear. MCG--6-30-15 is, however, known to possess complex, variable warm absorbers, which are mildly ionised outflows with velocities around 1000\kmps \citep{marinucci+2014,kammoun+2017}, in addition to evidence in the spectral variability for an ultrafast outflow \citep{igo+2020}. We can see in Fig.~\ref{fig:spectra:mcg} that the greatest difference in these flux spectra is the presence of the emission line-like feature at 6.8\keV. This feature is strongest in time intervals when the iron K and continuum band display a low degree of coherence when the feature is clearly separated from the broad iron K line by a dip in the spectrum at 6.7\keV, which may be an absorption feature, or simply a separation of the two emission features. In the time intervals in which the iron K band and continuum are coherent, but the time delay associated with reverberation is not seen (\textit{i.e.} the iron K band leads the continuum band), this 6.7\keV\ line feature is present, but blends into the blueshifted edge of the broad iron line, with no dip or absorption feature in between. This 6.8\keV\ feature is not present in the time intervals in which a reverberation-like time delay is detected (\textit{i.e.} the iron K band lags behind the continuum band), however a weaker line-like feature is seen at a higher energy of 7.05\keV. These features are likely associated with atomic transitions from ionised iron in the complex system of outflows.

\subsubsection{The variable part of the spectrum}

In addition to studying the X-ray flux spectrum (which is essentially a time-average over the selected time intervals), we may also use the wavelet covariance spectrum to obtain the shape of the variable part of the spectrum during the different time intervals, to understand which of the spectral components are contributing to the variability on different timescales and at different times. Just as for the time-resolved wavelet lag-energy spectra, we can compute the wavelet covariance spectra, filtering time periods based upon the coherence and relative phase between the iron K and continuum bands. The wavelet covariance spectra for IRAS\,13224$-$3809 and MCG--6-30-15 were computed from light curves extracted in 50 logarithmically-spaced energy bands between 0.3 and 10\keV.

\begin{figure*}
\centering
\subfigure[IRAS\,13224$-$3809] {
\includegraphics[width=0.49\textwidth]{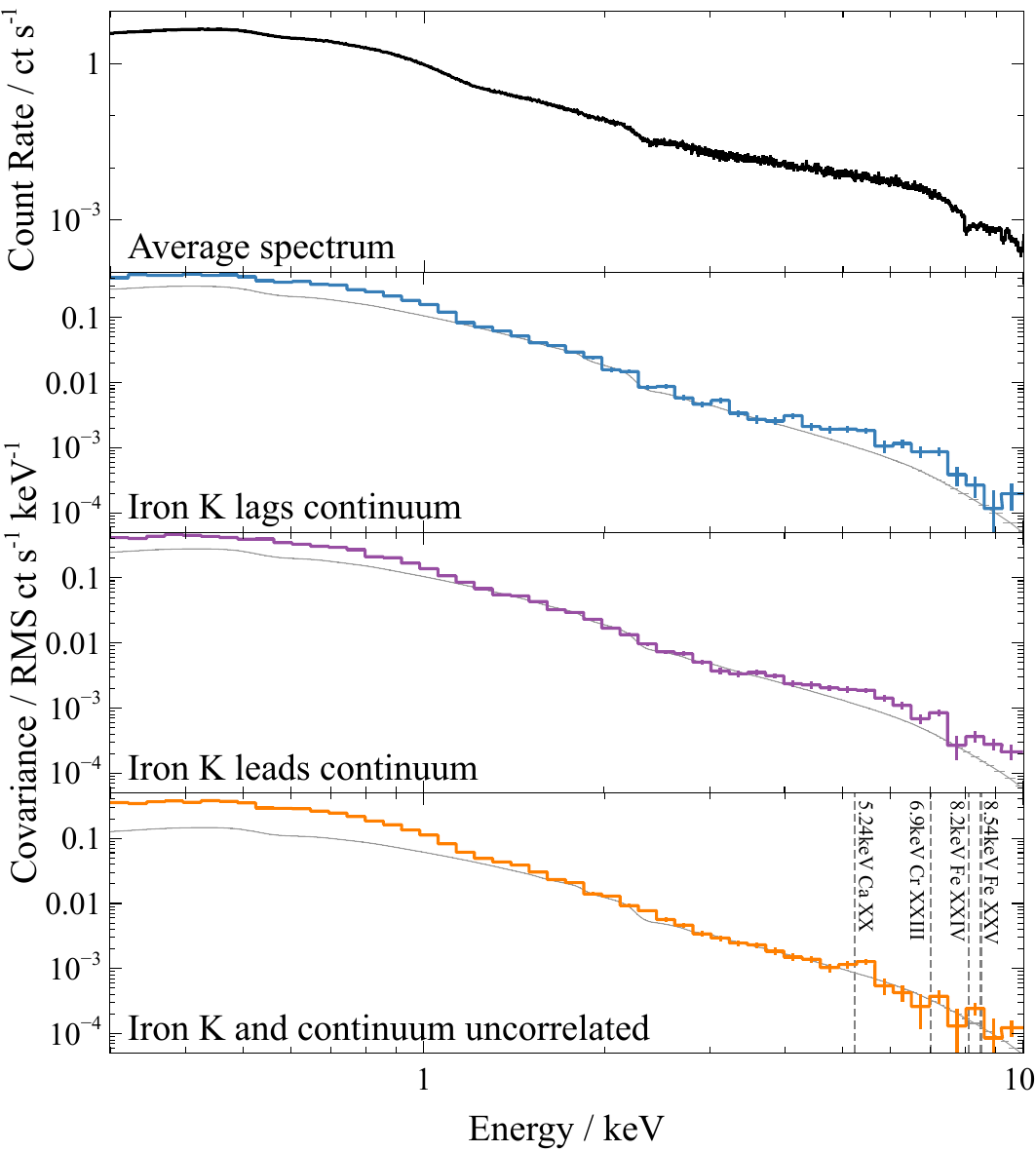}
\label{fig:covariance:iras}
}
\subfigure[MCG--6-30-15] {
\includegraphics[width=0.49\textwidth]{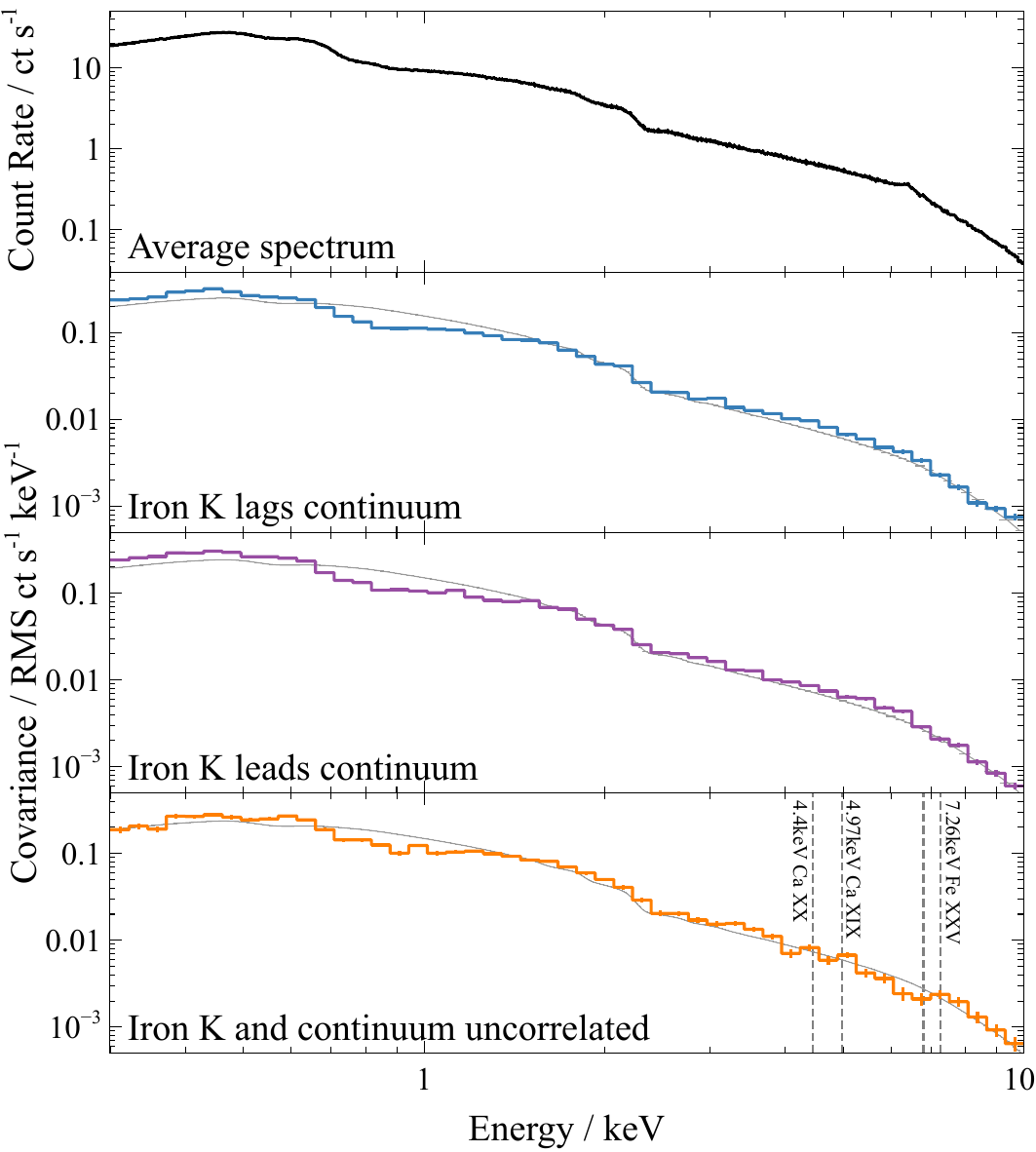}
\label{fig:covariance:mcg}
}
\caption[]{The covariance spectra for \subref{fig:covariance:iras} IRAS\,13224$-$3809 and \subref{fig:covariance:mcg} MCG--6-30-15, showing the variable part of the spectrum that is responding to variations on the frequency range in which reverberation is detected, derived from a wavelet analysis and separating time intervals in which (1) the iron K band lags behind the continuum and reverberation is detected in the iron K band, (2) the iron K band leads the continuum, and (3) the coherence between the two bands is low. The covariance spectra are compared to the shape of the time-average X-ray flux spectrum, shown in the top panels. Grey lines show the shape of the best-fitting power law continuum component. Vertical dashed lines indicate line-like features showing excess variance in specific atomic emission lines, identified by \citet{igo+2020} and attributed to variations of the ultrafast outflows in these AGN. In IRAS\,13224$-$3809, the atomic features are blueshifted according to an outflow velocity of $0.238c$, and in MCG--6-30-15, according to a velocity of $0.08c$.}
\label{fig:covariance}
\end{figure*}

Fig.~\ref{fig:covariance} shows the covariance spectra, obtained via wavelet analysis, from those same time periods. We see that during the time periods in which the iron K band and continuum band are coherent (both when variability in the iron K band lags and leads the variability in the continuum band) that the covariance spectrum represents the shape of the time-average flux spectrum. Relative to the best-fitting power law (which represents the primary X-ray continuum in the 0.3-10\keV\ band), excesses can be seen below 1\keV\ (the soft excess), and between 4 and 7\keV, corresponding to the shape of the relativistically broadened iron K line. In MCG--6-30-15, the effects of the warm absorber can also be seen in the time-averaged spectrum, leading to a flux decrement below the power law around 1\keV.

The correspondence between the covariance spectrum and the time-averaged spectrum during the periods when the iron K and continuum bands are coherent shows that the dominant mode of variability during these time periods is a change in the overall normalisation of the spectrum. Both the observed primary continuum and the reflection from the accretion disc are responding to changes in the luminosity of the corona. The shape of the observed spectrum (and the shape of the covariance spectrum) is modified by absorption along the line of sight, but the shape of spectrum remains approximately consistent as the transmission of the absorbing material is not changing substantially as a function of energy during these time periods.

On the other hand, the wavelet covariance spectrum in the time periods when the iron K and continuum bands are incoherent, does not show the relativistically broadened iron K line. During these time periods, the covariance spectrum in the iron K band follows the underlying power law continuum, with narrow line-like features of excess variability above the power law. These line-like features are similar to those observed in the excess variance spectra, found in some radio quiet AGN by \citet{igo+2020}. Unlike the wavelet covariance spectra, excess variance spectra show the variability integrated across all timescales longer than the binning time that is used, and averaged across the whole period of the observations, rather than being resolved into specific variability frequencies or timescales (\textit{c.f.} the Fourier or wavelet covariance spectrum) or resolved as a function of time (\textit{c.f.} the wavelet-derived spectrum). The lines correspond to the atomic transitions expected from an ionised outflow, with a consistent velocity shift applied to each of the lines, and are interpreted as showing the variability of an ultrafast outflow, a wind launched at relativistic velocity from the inner accretion disc. It should be noted that even though these are \textit{absorption} lines in the spectrum, they are seen as excess variability in the variance spectra as the variability in the emission lines is \textit{added} to the variability of the underlying continuum.

In the wavelet covariance spectra from the incoherent time periods, we observe the same set of lines that are seen in the excess variance spectra \citep{igo+2020}. In IRAS\,13224$-$3809, the lines are blueshifted corresponding to a velocity of $0.238c$. In the wavelet covariance spectrum, we find the same lines corresponding to Fe\,\textsc{xxv} and Ca\,\textsc{xx} transitions. We find an additional line at approximately 6.9\keV\ in the wavelet covariance spectrum. Assuming this arises from the same outflow component, this would correspond to the Cr\,\textsc{xxiii} transition that is expected to appear in this same energy band. We also note that these same line-like features begin to appear on top of the profile of the broad iron K line in the covariance spectrum in the time intervals in which the time lag is negative (\textit{i.e.} variability in the iron K band is leading that in the continuum band), suggesting that this inversion of the time lag occurs once the variability of the outflow starts to dominate, even though the fluxes in the iron K and continuum band are still linearly related (hence the coherence remains high).

In MCG--6-30-15, the outflow inferred from the lines in the excess variance spectrum is much slower, with the blueshift corresponding to a velocity of $0.08c$. We find the same features that correspond to Fe\,\textsc{xxv} and Ca\,\textsc{xx} transitions, in addition to transition that is not apparent in the time-averaged excess variance spectrum, that we can attribute to Ca\,\textsc{xix} at around 5\keV.

From the combination of the flux spectra and covariance spectra we can understand why the iron K reverberation signal from the inner accretion disc is transitory in nature. We find that there are time intervals in which the variability is dominated by normalisation changes of the overall spectrum, during which times we see the iron K line from the inner accretion disc responding to changes in luminosity of the corona and measure the corresponding reverberation lag. On the other hand, the iron K reverberation signal from the inner accretion disc is not detected during time periods in which absorption from outflows is more pronounced in the spectrum. Changes in the properties of the outflow are dominating the variability that we observe iron K band, and even though the inner disc may still be illuminated by the corona, we are unable to detect the corresponding reverberation signal at these times.

\section{Time evolution of the corona}
\label{sec:evolution}
The ability of wavelet spectral timing analysis to trace the variation in lag as a function of both frequency and time allows us to study how the reverberation timescale between the corona and accretion disc evolves. The X-ray emission from the corona is highly variable. Tracing how the reverberation time lag correlates with the X-ray luminosity and other properties of the corona will give us important insight into how changes in the geometric properties of the corona (\textit{e.g.} its location and size) are related to the variability we observe, shedding light on the process by which the corona is formed and energised from the accretion flow.

While we find that the reverberation signal in the iron K band is transitory in nature, and can disappear during time intervals in which the variability is dominated by ultrafast outflows, the reverberation signal in the soft X-ray (0.3-1\keV) band is much more stable. Where reverberation is observed from the inner accretion disc, the soft X-ray band is expected to be dominated by the \textit{soft excess}, formed from the accretion disc where relativistic effects cause a number of soft X-ray emission lines to be blended together. The stability of the soft X-ray reverberation signal, and the higher signal-to-noise in the soft X-ray band compared to the iron K band, means that we may apply wavelet spectral timing analysis between the soft X-ray (0.3-1\keV) and continuum (1.2-4\keV) bands to trace the evolution of the corona in time.

It is possible that other emission components contribute to the soft excess, for example Comptonisation in a warm atmosphere above the disc \citep[\textit{e.g.}][]{done_jin}, and the shape of the soft X-ray spectrum can be modified by warm absorbers (moderately ionised outflows from the AGN), and each of these components will modify the variability and lag spectra over the frequency ranges in which they operate \citep[\textit{e.g.}][]{silva+2016}. Performing a frequency-resolved spectral timing analysis probes only those components that are responding to variability on the chosen timescale, thus we may separate out reverberation from the inner disc by selecting the frequency range over which both the iron K and soft excess are detected to respond to continuum variations.

\subsection{The evolution of the corona during an X-ray flare}
A particularly dramatic example of variability in the X-ray emitting corona around a supermassive black hole was observed in the AGN I Zwicky 1 (1\,Zw\,1). In 2020 January, a series of bright, short-duration X-ray flares were observed by \textit{XMM-Newton} and \textit{NuSTAR}, during which the count rate increased by a factor of around 2.5 for periods of around 10\ks\ \citep{1zw1_nature}. During the flares, the X-ray spectrum showed a decrease in the high-energy cut-off and steepening of the continuum spectrum, accompanied by a drop in the strength of the reflection observed from the accretion disc relative to the continuum (\textit{i.e.} the reflection fraction). These observations were interpreted as the cooling and the acceleration of the corona away from the black hole and accretion disc during the flares \citep{1zw1_flare_paper}.

\begin{figure}
\centering
\includegraphics[width=0.48\textwidth]{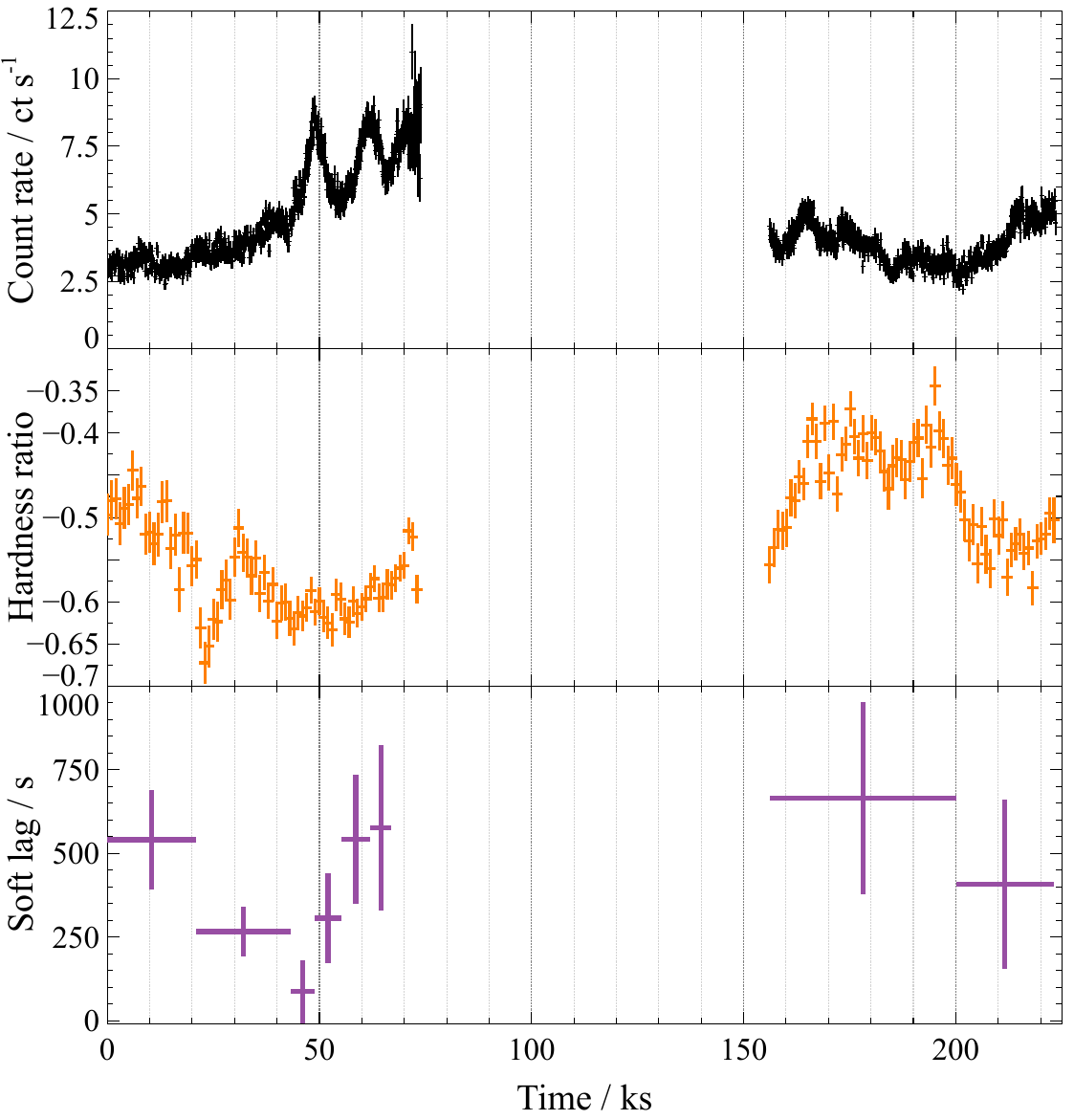}
\caption[]{Evolution of the soft X-ray reverberation lag during X-ray flares observed from the AN I\,Zw\,1 in 2020. The top panel shows the 0.3-10\keV\ light curve. The middle panel shows the hardness ratio, defined between the 2-10\keV\ hard band and 0.3-1\keV\ soft band as $(H-S)/(H+S)$. The lower panel shows the time lag, measured using by wavelet analysis, averaged into time bins and selecting the frequency with the greatest average lag amplitude in each bin.}
\label{fig:1zw1_flare_lag}
\end{figure}

Calculating the time lag between variations in the soft X-ray and continuum bands as a function of time over the course of these observations using wavelet analysis reveals how the reverberation timescale between the corona and disc changes during the flares (Fig.~\ref{fig:1zw1_flare_lag}). To enhance the signal-to-noise during the short time periods over the course of the flare, this wavelet analysis was conducted on the light curves summed from both the EPIC pn and MOS detectors. The time lag as a function of frequency (wavelet scale) was averaged into a series of time bins, then in each of the time bins, the frequency with the greatest lag amplitude (the most negative lag from Fig.~\ref{fig:lagfreq}) was selected to define the reverberation time lag (as was done in \citealt{demarco+2012} and \citealt{kara_global}). Selecting the frequency with the greatest lag amplitude allows us to account for the fact that the frequency or timescale at which reverberation is observed may vary \citep{alston_iras2}. The time bins correspond to the rising and falling halves of each flare, in addition to two bins both before and after the flares. The time periods before and after the flares were divided into two time bins based on observed changes in the hardness ratio of the X-ray emission \citep{1zw1_flare_paper}.

Before the flares begin, we see a steady decline in the reverberation timescale, which may be interpreted as a contraction of the corona to a more confined region around the black hole. Strictly, the reverberation time is most closely related to the vertical scale height of the corona above the plane of the disc \citep{lag_spectra_paper}, however simultaneous spectral and timing analysis of I\,Zw\,1 requires that the corona is also confined radially to a small region around the black hole \citep{1zw1_corona_paper}.

During the X-ray flares, we see a rapid increase in the reverberation time between the corona and disc, corresponding to a rapid expansion of the corona. The scale height of the corona above the disc increases rapidly during both the rise and fall of the first flare, then remains large during the second flare. The soft X-ray lag time increases from an upper limit of 180\s\ as the flares begin, to $(540\pm 190)$\s\ during the second flare. Assuming a black hole mass of $3\times 10^{7}$\Msun\ \citep{vestergaard+06, 1zw1_nature} these raw time lags correspond to the (uncorrected) light travel time over a distance increasing from  an upper limit of 1.2\rg\ to $4\pm 1$\rg.

In reality, we do not measure energy bands that consist of solely continuum emission, or solely reflected emission from the accretion disc. Rather our continuum-dominated band will have some contribution of the delayed reflected emission and our reflection band will have some contribution of promptly-responding continuum emission. This results in the measured time lag being \textit{diluted} from the intrinsic light travel time delay \citep{understanding_emis_paper,cackett_ngc4151}, where the `intrinsic' delay is the integrated average through the impulse response function, accounting for all possible light paths from the corona to the disc. If our `reflection' band is a sum of reflected flux $R_1$ and continuum flux $C_1$, and the ratio of reflected to continuum flux in this band is $F_\mathrm{ref} = R_1 / C_1$, and likewise the ratio of reflected to continuum fluxes, $R_2$ and $C_2$, in our `continuum' band is $F_\mathrm{cont} = R_2 / C_2$, we may define the dilution factor $D$, such that the measured time lag, $\tau$, is $D$ times the intrinsic time lag, $\tau_0$. Applying a small-angle approximation (valid in the limit of low frequencies, where $\nu \ll 1/2\pi\tau$):
\begin{align}
\label{equ:dilution}
D \approx \frac{F_\mathrm{ref}}{1+F_\mathrm{ref}} - \frac{F_\mathrm{cont}}{1+F_\mathrm{cont}}
\end{align}

Using the best-fitting model to the X-ray spectrum of I\,Zw\,1 \citep{1zw1_flare_paper} to estimate the fraction of the photon flux in the 0.3-1\keV\ and 1.2-4\keV\ bands that is contributed by the reflection and continuum components, we can estimate that the dilution factor for the soft X-ray lags in this AGN is $D\approx 0.1$. From the measured soft X-ray lags, we can therefore estimate that the light travel time between the corona and disc is increasing from $\la 12\,r_\mathrm{g}/c$ to $40\pm 10\,r_\mathrm{g}/c$ as the flares are launched.

To estimate the scale height of the corona from the average light travel time between the corona and disc, one must account for the time delays experienced by rays as they propagate through the strong gravitational field close to the black hole \citep{shapiro}, and average over all possible light paths from the corona to the disc to the observer (see Appendix~\ref{app:raytrace}). Approximating the corona as a point source located on the spin axis of the black hole, and using a General Relativistic ray tracing code \citep{understanding_emis_paper,propagating_lag_paper}, we can estimate that the characteristic scale height of the corona, from which we may consider the bulk of the X-ray emission from originating, increases from $\la 2$\rg\ to $(12\pm 4)$\rg\ above the plane of the disc as the flares are launched.

\subsection{Stochastic variability of the corona}
We may also use wavelet spectral timing analysis to trace the continuous variability of the corona that gives rise to the stochastic variations observed in the X-ray emission of IRAS\,13224$-$3809. Following the same methodology used to trace the evolution during the flare in I\,Zw\,1, the wavelet coherence and lag spectra were calculated between the 1.2-4\keV\ continuum and 0.3-1\keV\ soft X-ray bands for each of the observations and the lag \textit{vs.} frequency or wavelet scale was averaged into 1000\s\ time bins. Once again, the frequency in each time bin with the greatest lag amplitude was taken to represent the soft X-ray lag in that bin.

\begin{figure*}
\centering
\includegraphics[width=0.95\textwidth]{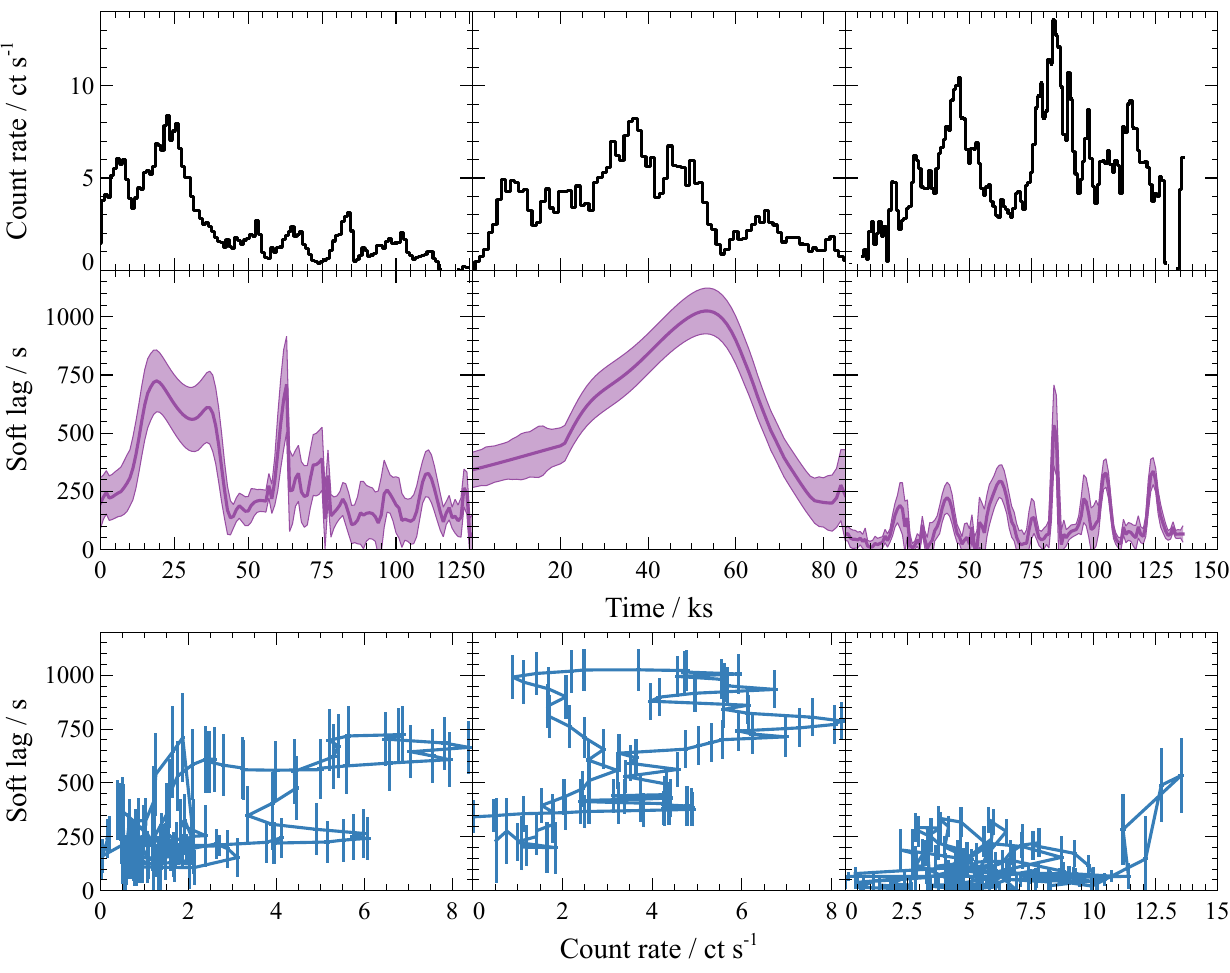}
\caption[]{The variation of the soft X-ray reverberation lag during three observations of IRAS\,13224$-$3809, computed between the 1.2-4\keV\ continuum band and the 0.3-1\keV\ band dominated by the soft X-ray emission reprocessed by the disc, taking the maximum lag amplitude over the $3\times 10^{-4}$ to $2\times 10^{-3}$\Hz\ frequency range. The top panel shows the X-ray light curve over the full 0.3-10\keV\ band and the second panel shows the soft lag, computed via wavelet analysis of the light curves, both in 1000\s\ time bins. The lower panels show the soft lag plotted \textit{vs.} the count rate for each of those three observations.}
\label{fig:softlag_time}
\end{figure*}

Fig.~\ref{fig:softlag_time} shows the light curves and the soft X-ray lag as a function of time during three of the \textit{XMM-Newton} observations of IRAS\,13224$-$3809, in addition to the time lag \textit{vs.} the 0.3-10\keV\ count rate. In the best-fitting model to the X-ray spectrum of IRAS\,13224$-$3809, the soft X-ray band is dominated by reflection from the accretion disc \citep{jiang_iras}, hence the soft X-ray lag can once again be interpreted in the context of the scale height of the corona above the plane of the disc \citep{alston_iras2, caballero-garcia+2020}.

We see that there is not a simple correspondence between the X-ray luminosity and the soft X-ray time lag, and that the size or scale height of the corona does not simply increase in tandem with rising X-ray luminosity. There are time periods in which the reverberation time lag varies erratically (in a similar manner to the count rate), and there are time periods in which the reverberation lag evolves smoothly, even though the count rate is erratic.

The time series of the soft X-ray lag reveal, however, that the most significant peaks in the light curve are, indeed, accompanied by an expansion of the corona, just not always at the same time. In the left panel of Fig.~\ref{fig:softlag_time} we can see that the peak in the light curve at the beginning of the observation corresponds to a period of longer soft X-ray lags and hence a greater scale height of the corona. In the middle panel, however, we see that the smooth rise and fall of the reverberation time lag (detected at $6\sigma$ significance) is delayed with respect to the peak at the beginning of the light curve.

In the right panel, we see an example of a time period in which both the X-ray luminosity and the reverberation lag are erratic. There are a series of three short peaks as the count rate climbs at the beginning of the observation, each accompanied by an increase in the reverberation lag (and an expansion of the corona). Each of the peaks in the reverberation lag, however, is delayed with respect to the peaks in the light curve by a time period increasing from 5000\s\ to 15,000\s. The greatest peak in the light curve, approximately 84\ks\ into the observation, is also accompanied by an increase in the reverberation lag time, though the increase in reverberation lag is much shorter-lived than the peak in the light curve. While the peak in the light curve lasts approximately 20\ks, the peak in the reverberation lag last only 7000\s\, and begins 10\ks\ after the rise in count rate. Each of the short-lived increases in reverberation lag is detected at approximately $3\sigma$ significance in 1000\s\ time bins.

Plotting the soft X-ray lag against the count rate (shown in the lower panels of Fig.~\ref{fig:softlag_time}) reveals a hysteresis in the behaviour of the corona. In each case, we see that during the peaks in X-ray luminosity, the corona traces a counter-clockwise loop in the reverberation lag \textit{vs.} count rate plane. The corona initially brightens at constant scale height, then the corona expands, then the luminosity drops in most cases before the corona contracts (except for the sharp peak in the right panel, where the expansion of the corona is extremely short-lived).

\begin{figure}
\centering
\includegraphics[width=0.48\textwidth]{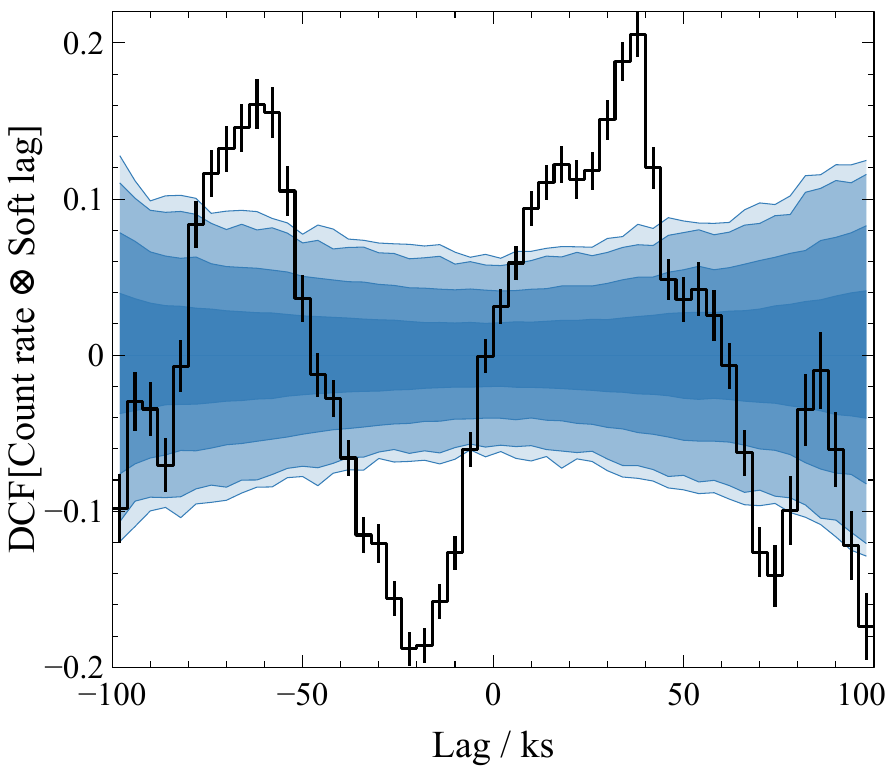}
\caption[]{The discrete correlation function (DCF) between the X-ray count rate and the soft X-ray reverberation lags, averaged over the 2.2\Ms\ total observations of IRAS\,13224$-$3809. The shaded regions shows the 1, 2, 3 and $4\sigma$ confidence intervals for the case where the time series were uncorrelated.}
\label{fig:dcf}
\end{figure}

To study the correlation between the soft X-ray lag and the luminosity of the corona, we may calculate the cross-correlation function between the time lag and count rate using the discrete correlation function (DCF) to compute the correlation over the entire 1.5\Ms\ of discontinuous observations \citep{edelson_fvar}. Fig~\ref{fig:dcf} shows the DCF between the soft X-ray lag and total 0.3-10\keV\ count rate in IRAS\,13224$-$3809, compared to confidence limits of the DCF if the two time series were uncorrelated (estimated by randomising the ordering of the data points to produce two uncorrelated series with the same distribution as the observed lag and count rate). A correlation is detected between the X-ray count rate and the soft X-ray lag at greater than $5\sigma$ significance. The expansion of the corona is found to lag behind the increase in count rate, with the DCF peaking at a time lag of $38$\ks\ (with a broad range of lags between 10 and 40\ks). We also note a periodic behaviour of the DCF that occurs due to repeated short-timescale brightening events that can be seen in the light curve. They occur on average every approximately 100\ks\ and give rise to a secondary (albeit weaker) peak in the DCF at a lag of $-62$\ks, where a peak in count rate can be associated with the peak in reverberation lag from the previous brightening event. We also note a significant anticorrelation at a lag of $-20$\ks\ suggesting that the corona tends to contract in size before each brightening event.

In IRAS\,13224$-$3809, the soft X-ray reverberation lag varies from a baseline level of around 100$\sim$150\s\ to peaks as high as 300$\sim$1000\s. Assuming a black hole mass of $(6_{-4}^{+12} \times 10^{6})$\Msun\ in IRAS\,13224$-$3809 \citep{gonzalezmartin+2012,alston_iras2}, this corresponds to the light crossing time over a distance increasing from 3$\sim$5\rg\ to 10$\sim$30\rg\ during the peaks. From the best-fitting model to the X-ray spectrum \citep{jiang_iras}, we can estimate a soft lag dilution factor of $D\approx 0.4$ and infer that the light travel time from the corona to the disc in IRAS\,13224$-$3809 is increasing from a baseline of $8\sim 13\,r_\mathrm{g}/c$\ to $25\sim 75\,r_\mathrm{g}/c$ during the peaks in the X-ray luminosity. Approximating the corona as a point source close to the black hole, and applying the General Relativistic ray tracing code, we estimate that the scale height of the corona increases from $< 2$\rg\ at the baseline to $6\sim 18$\rg\ at the peaks (see Appendix~\ref{app:raytrace}). No correlation is observed between the soft X-ray lag and the ratio of the fluxes in the soft X-ray and continuum bands, and the range of dilution factors that are calculated from the high and low flux spectra is insufficient to account for the range of lag times that are observed. This shows that variation of the scale height of the corona is required, and the changing lag time cannot be attributed to changes in other spectral components that change the dilution factor.

\section{Discussion}
Performing X-ray spectral timing using wavelet analysis and wavelet transforms allows us to trace the time variability of the timing properties of the emission from accreting systems (including AGN, stellar mass black holes and neutron stars). A Fourier transform describes the variability as a function of Fourier frequency or timescale, but averaged over the duration of the light curve segment for which it is computed, implicitly assuming stationarity of the process that generates the observed variability. A wavelet transform is analogous to a Fourier transform, but describing the variability as a function of frequency/timescale, and its variation in time.

In principle, wavelet analogues can be defined for any of the standard spectral timing products, including the power spectrum or periodogram, the cross spectrum, the coherence and the time lag spectrum \citep{reverb_review}, simply by replacing the Fourier transform for the wavelet transform. These wavelet spectral timing products then trace the spectral timing information at each point in time. Some care must be taken, since the most general form of the wavelet transform (the continuous wavelet transform, which best-traces the evolution of frequency components in time and provides the most intuitive description of the signal) does not strictly provide a unique composition of the signal. This means that while we can write down approximate expressions for the uncertainties or error bars, these expressions are only approximate and should be confirmed using Monte Carlo simulations or similar.

Here, we have focused on using wavelet spectral timing to trace the time variability of X-ray reverberation signals from the inner accretion disc. Phase lags are measured between the light curves in different energy bands in order to probe the time delay as variations in the luminosity of the X-ray continuum emitted from the corona propagate to the accretion disc and are subsequently observed in the reprocessed emission, either in the soft X-ray band, or the relativistically broadened iron K line \citep{fabian+09}. The reverberation time delay is interpreted in the context of the light travel time between the corona and accretion disc and can be used to constrain the location and geometry of the corona, specifically, the scale height of the corona above the disc \citep{lag_spectra_paper, cackett_ngc4151, propagating_lag_paper}.

\subsection{X-ray reverberation hidden beneath variable outflows}
Wavelet spectral timing analyses of the AGN IRAS\,13224$-$3809 and MCG--6-30-15 reveal the variable and transitory nature of the reverberation signal in the iron K band. There are time periods in which reverberation is detected in the iron K band, namely when the coherence between the light curves in the iron K and continuum bands is high (the light curves show a high degree of correlation), and variability in the iron K band lags behind that in the continuum band. There are, however, also significant time periods when reverberation is not detected, when either variability in the iron K band leads that in the continuum, or the iron K and continuum bands show a low degree of correlation. On the other hand, the reverberation signal in the soft X-ray band is much more persistent in time (although the amplitude of the soft X-ray reverberation lag is variable).

We find that the average time lag \textit{vs.} energy spectrum, taking just the time periods in which the iron K lag is observed (Fig.~\ref{fig:wavelet_lagen_reverb}), resembles the form that is commonly attributed to X-ray reverberation from the inner accretion disc among Seyfert 1 AGN \citep{kara+13,kara_global}. The earliest response is seen in the energy range most strongly dominated by continuum emission observed directly from the corona, between 1 and 3\keV. The energy bands dominated by reprocessed emission from the accretion disc respond later, namely the iron K line in the 4-7\keV\ band and the 0.3-1\keV\ soft X-ray band, where emission lines from the disc are blended together as they are broadened by Doppler shifts and gravitational redshifts. The redshifted wing of the iron K line, between 3 and 6\keV\ is comprised of gravitationally redshifted photons from the innermost radii on the accretion disc, closest to the corona, so is seen to respond at earlier times than the 6\keV\ iron K photons from larger radii \citep{zoghbi+2012}.

\begin{figure}
\centering
\includegraphics[width=0.48\textwidth]{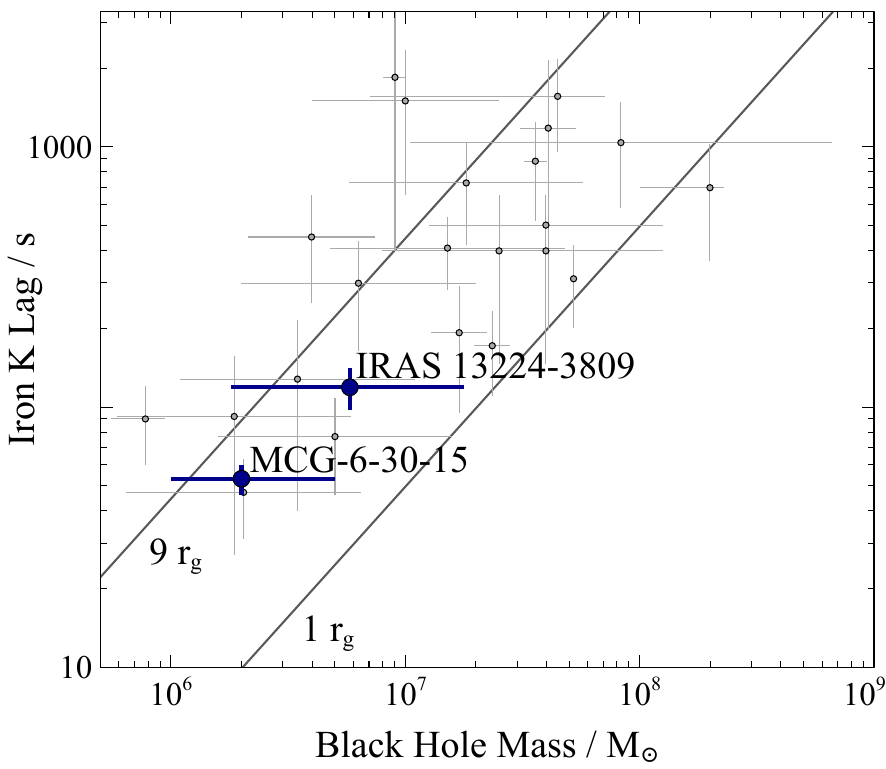}
\caption[]{The iron K reverberation lag times measured in IRAS\,13224$-$3809 and MCG--6-30-15 using wavelet analysis, \textit{vs.} the black hole mass, compared to the lag times in the sample of Seyfert-type AGN in which reverberation has been detected, from \citet{kara_global}, which were measured using conventional Fourier analysis techniques. Lines show the light crossing times across distances corresponding to 1 and 9\rg\ for the respective black hole masses.}
\label{fig:lag_mass}
\end{figure}

Defining the characteristic iron K reverberation time as the difference in the lag-energy spectrum between the earliest responding bin in the 1-3\keV\ continuum band and the peak of the iron K line, we can compare the iron K reverberation lag as a function of black hole mass to the rest of the sample of Seyfert galaxies in which reverberation has been detected \citep{kara_global} in Fig.~\ref{fig:lag_mass}. The reverberation lag time correlates strongly with the mass of the black hole and corresponds to the light crossing time across just $1\sim 10$\rg\ (where the gravitational radius is the coordinate size of the event horizon of a maximally spinning black hole). The iron K reverberation lags measured using wavelet analysis in IRAS\,13224$-$3809 and MCG--6-30-15 follow this same relation, showing that wavelet spectral timing is able to extract the equivalent reverberation signal. Wavelet analysis is able to separate the signal in time from other mechanisms of variability that mask reverberation when time-averaged measurements are made using conventional spectral timing techniques. While Fourier analysis can separate different mechanisms of the variability based upon the timescales or frequencies at which they operate, wavelet analysis can separate mechanisms of variability by both frequency and time.

Comparing the X-ray flux spectrum from the time periods during which reverberation is and is not detected in IRAS\,13224$-$3809 and MCG--6-30-15, as well as the covariance spectra, which show the part of the spectrum that is varying on the relevant timescale, we can understand what is causing iron K reverberation to appear and disappear in these two AGN. During the time periods in which reverberation is detected, the covariance spectrum follows the same form as the time-averaged X-ray flux spectrum, showing that during these time periods, the variability is largely described by an overall change in the normalisation of the spectrum (\textit{i.e.} all of the spectral components are varying in tandem with a change in the continuum luminosity).

In IRAS\,13224$-$3809, we see that during the time periods in which variability in the iron K band is leading that in the continuum-dominated band, absorption from the ultrafast outflow (UFO) becomes stronger. This outflow, launched at a velocity of $\sim0.24c$ from the inner accretion disc, is detected via an Fe\,\textsc{xxv} absorption line imprinted in the spectrum around 8\keV\ that is observed to vary. The lag \textit{vs.} energy spectrum during these time periods shows that the variability at lower energy X-rays systematically lags behind that at high energies, in contrast to the reverberation where variability in both the soft X-ray and iron K bands lags behind a continuum band in between. \citet{parker_iras_nature} show that the UFO in IRAS\,13224$-$3809 is highly variable, and the absorption lines appear most strongly in the spectrum during the periods of lowest X-ray luminosity.

\citet{silva+2016} show that such soft X-ray lags can appear when the ionisation state of the absorbing gas is changing in response to changes in the continuum luminosity. Variations in the ionisation parameter lag behind the variations in the ionising luminosity according to the recombination time within the gas. The ionisation changes then translate into changes in the continuum transmission at lower energies, with lower ionisation gas absorbing more of the continuum), hence producing the lag in the soft energies. \citet{silva+2016} model the variability of \textit{warm absorbers}, relatively low velocity, moderately ionised gas, at relatively large distances from the black hole. As such, the soft X-ray lags related to ionisation changes of the warm absorbers are seen just on the longest variability timescales or lowest Fourier frequencies. In the case of IRAS\,13224$-$3809, we are seeing these soft X-ray lags at the much higher Fourier frequencies at which we observe reverberation from the inner disc. We attribute this behaviour not to warm absorbers, but to the UFO that is launched from the inner accretion disc, hence the variability is seen on much shorter timescales.

In both IRAS\,13224$-$3809 and MCG--6-30-15, we find that there are also periods when low coherence is measured between variability in the iron K and continuum bands, amounting to 19 per cent of the time in IRAS\,13224$-$3809 and 43 per cent of the time in MCG--6-30-15. In the time-averaged X-ray flux spectra, we see that absorption features are the strongest during these periods. The covariance spectra in the iron K band at these times do not resemble the time-average spectra, but rather take the form of a power law with narrow line features similar to those reported in the excess variance spectra by \citet{igo+2020}. These features correspond to blueshifted absorption lines expected from the highly-ionised gas that are expected from the UFO and appear in the spectrum when the variability is dominated by changes to the emission lines. Via the wavelet covariance spectra, we discover that the degree of variability of the UFO absorption lines changes over time. The low coherence between the continuum and the iron K band shows that during the time periods in which the UFO lines are most variable, there is not a simple linear relationship between the variability in the continuum and iron K band. Changes to the spectral shape, specifically changes to the narrow absorption lines, introduce a non-linearity in the response of the system to variations in the X-ray continuum.

Reverberation is not observed in the iron K line from the inner disc during time periods in which there is significant variability in the ultrafast outflows. This is not to say that the inner accretion disc is not illuminated by the corona at these times, rather these findings suggest that the variation induced by changes to the properties of the UFOs is so significant that it dominates the variability of the iron K band. In other Seyfert type AGN, ultrafast outflows are either not present, or do not seem to dominate the variability for such large fractions of the time. We find in IRAS\,13224$-$3809 that reverberation from the inner disc is observable in the iron K band for approximately 49 per cent of the time, while UFO variability dominates for the remaining 51 per cent of the time. In MCG--6-30-15, reverberation is observable in the iron K band only 23 per cent of the time.

The best-fitting model to the broad iron K line in the spectrum of IRAS\,13224$-$3809 indicates that we are observing the accretion disc at a relatively high inclination angle, from 45\,deg up to as high as 80\,deg to the axis \citep{caballero-garcia+2020,jiang_iras}. If the UFO is a magnetically-driven outflow from the inner accretion disc, material should be ejected by centrifugal forces along field outward-pointing magnetic field lines making an angle of less than 60\,deg to the surface of the disc \citep{blandford_payne}. Therefore lines of sight at inclination angles more than 30 degrees from the axis will intersect the outflow and observe a high degree of variability from the UFO, and the greater the inclination, the greater the more of the outflow is travelling in the direction of the observer. This means that highly variable UFOs are more likely to be observed from accretion discs observed closer to edge-on, and at high inclinations, the most variable UFOs are better able to mask the iron K reverberation signal from the disc. 

In MCG--6-30-15, the disc is observed at an inclination of only around 30\,deg \citep{marinucci+2014,caballero-garcia+2018} and the outflow is measured at a much lower velocity in this AGN of only around $0.08c$. The observed velocity of the outflow will be the projection of the intrinsic velocity along the line of sight, hence the lower measured velocity is consistent with the lower inclination. For the outflow to explain the masking of the iron K reverberation signal in this case, it is likely that in MCG--6-30-15 either it is intrinsically stronger, with a greater degree of variability on timescales equal to the reverberation timescale, or it is launched across a greater range of angles from the disc surface.

\subsection{The variable corona and the launching of X-ray flares}
A picture is emerging of how variations in the luminosity of X-ray emitting coron\ae\ around black holes are connected to changes in its location and geometry. The spatial extent of the corona can be inferred from the pattern of illumination it produces on the accretion disc, measured via the profile of the relativistically broadened iron K line. During the highest X-ray flux states, the corona is found to expand to a larger region over the innermost parts of the accretion disc, extending a few tens of gravitational radii over the inner disc \citep{mrk335_corona_paper, 1h0707_var_paper}, while during low X-ray flux states, the corona collapses to a confined region within just a few gravitational radii of the singularity \citep{1h0707_jan11,parker_mrk335,mrk335_corona_paper}. 

Additional evidence for the expansion of the corona during the high X-ray flux states has been found in measurements of the soft X-ray reverberation lag between different observations of IRAS\,13224$-$3809. The reverberation timescale between the corona and disc was found to increase between observations with increasing 2-10\keV\ X-ray luminosity \citep{alston_iras2, caballero-garcia+2020}. Each of these time lag measurements was averaged over observations lasting around 125\ks, a much longer timescale than many of the most extreme flux variations that are observed in the light curve. Wavelet spectral timing analysis enables us to track the variation of the reverberation lag over much shorter timescales, and to relate short-timescale variation of the lag time to the extreme, short timescale variability that is observed in the light curve, with X-ray count rates typically varying by factors of a few on timescales of just hours.

Some of the most extreme examples of X-ray variability in AGN come in the form of short-duration X-ray flares, where the count rate is observed to increase by factors of a few for periods of a few tens of kiloseconds \citep[\textit{e.g.}][]{1zw1_nature}, or even more extreme cases of flux increases of a factor of 10 during flares lasting around ten days \citep[\textit{e.g.}][]{mrk335_flare_paper}. Evolution of the X-ray spectrum during the flares, in particular the sudden drop in the strength of the reflection from the disc relative to the strength of the continuum (the reflection fraction) suggests that as the flares are launched, the corona is accelerated away from the black hole and accretion disc, with its temperature dropping in the process, before collapsing to a confined region around the black hole after the flare subsides \citep{mrk335_flare_paper, 1zw1_flare_paper}.

Using wavelet analysis to trace the variation in the soft X-ray reverberation lag provides further evidence for the expansion or upwards motion of the corona during these flares. The reverberation lag increases rapidly in tandem with the rise of the first of a series of flares observed from the Seyfert 1 AGN I\,Zw\,1 in 2020 January, then remains at the increased level during the subsequent flares. Accounting for spectral dilution of the time lags, the increasing reverberation lag is consistent with the corona increasing from a scale height of $\la 2$\rg\ to $(12\pm 4)$\rg\ above the disc as the flares are launched (assuming that the radial extent of the corona is small such that a point source approximation can be used to translate the measured time lag into a height). This measurement of the increase of the scale height falls alongside the observation that the corona accelerates to a velocity of $\sim 0.9$c away from the disc during the flares, the measurement of the temperature dropping from $140_{-20}^{+100}$\keV\ to $45_{-9}^{+40}$\keV\ and the tentative measurement of the radial extent of the corona over the inner disc, increasing from $7_{-2}^{+4}$\rg\ to $18_{-7}^{+6}$\rg\ \citep{1zw1_flare_paper}.

\subsection{The disc-corona connection}
The variability of the corona that accompanies the short timescale stochastic variability of the X-ray flux is more complicated. We observe that there is not a simple one-to-one mapping between the X-ray count rate and the reverberation lag on the shortest variability timescales. We do, however, observe that the trend of increasing lag time with increasing X-ray luminosity that was observed on long timescales, in general, holds true on the shortest timescales. There are time periods during which the reverberation lag varies smoothly over a relatively long timescale, and timescales when the corona appears to be varying rapidly, with short-lived peaks in the reverberation time lag tracking with peaks in the X-ray light curve lasting a few tens of kiloseconds. 

We observe a hysteresis in the behaviour of the corona during the brightest peaks (or `mini flares') in the light curve. The cross-correlation function between the X-ray count rate and reverberation lag time (Fig.~\ref{fig:dcf}) reveals a time delay of between 10 and 40\ks\ between the X-ray count rate increasing and the corresponding increase in the scale height of the corona. The correlation and time lag between the count rate and time lag are detected in IRAS\,13224$-$3809 at greater than $5\sigma$ significance.

We can interpret this time delay in the context of an X-ray continuum that is produced by from the Comptonisation of seed photons from the accretion disc in a compact corona close to the black hole (for which the simplest model is a point-like corona on the rotation axis above the black hole, sometimes referred to as a `lamppost'). Using general relativistic ray tracing simulations between the corona and accretion disc \citep[\textit{e.g.}][]{propagating_lag_paper}, we may show that approximately 80 per cent of the thermal photon flux from the accretion disc that passes through a small corona at a height of 5\rg\ above the black hole, originates from radii less than 7\rg\ on the accretion disc once gravitational light bending is accounted for, and 90 per cent from within 10\rg.

If the X-ray luminosity is modulated by accretion rate fluctuations propagating inwards through the accretion disc, a localised increase in accretion rate (\textit{e.g.} a `clump' in the disc) will first result in enhanced emission of thermal photons from the disc. This clump will begin significantly modulating the thermal seed photon flux passing through the corona once it reaches the inner $7\sim 10$\rg\ of the disc. Estimating the viscous propagation timescale through a standard accretion disc, for example using Equation 1 of \citet{mrk335_flare_paper}, and estimating an approximately flat, thin accretion disc with $(h / r)\sim r^{-1}$ (\textit{i.e.} $\beta = 2$), the viscous propagation time from $7\sim 10$\rg\ to the innermost stable circular orbit (ISCO) is $500\sim 2000\,GMc^{-3}$, which for the mass of the black hole in IRAS\,13224$-$3809, corresponds a propagation time of $15\sim 60$\ks, in approximate agreement with the $10\sim 40$\ks\ time delay that is observed. The seed photon flux through the corona increases as the clump moves through the inner disc, which leads to the increase in X-ray count rate. If the corona itself, however, is energised by magnetic fields anchored predominantly to the inner edge of the accretion disc, it is not until the clump reaches the inner edge of the disc that the structure of the corona is affected, leading to the time lag we observe between the increase in count rate and the increase in scale height.

\begin{figure*}
\centering
\subfigure[Point-like corona] {
\includegraphics[width=0.98\textwidth]{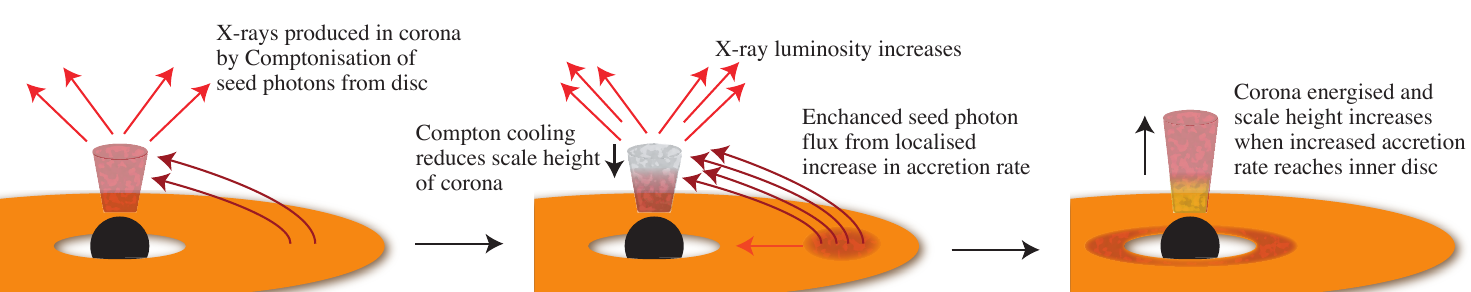}
\label{fig:disc_corona:point}
}
\subfigure[Two component corona, with extended component over disc and collimated core or jet base] {
\includegraphics[width=0.98\textwidth]{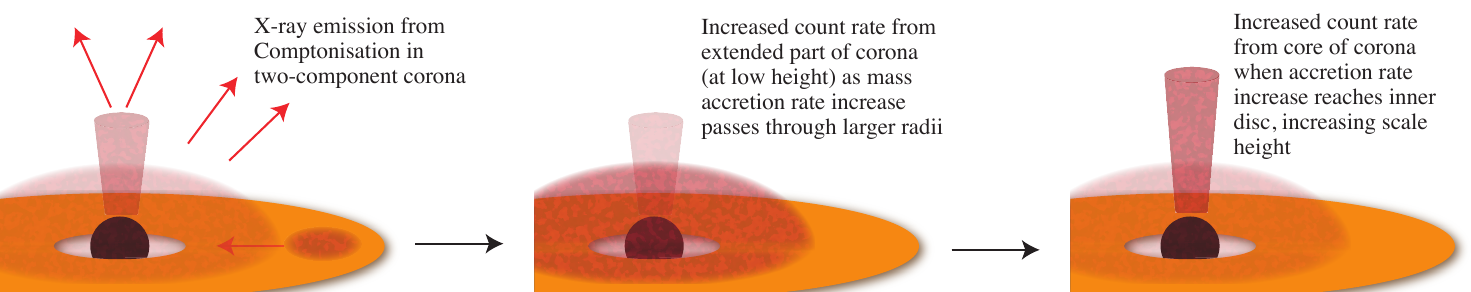}
\label{fig:disc_corona:ext}
}
\caption[]{Two possible scenarios for the observed relationship between the X-ray luminosity and the scale height of the corona, modulated by a localised increase in mass accretion rate (\textit{i.e.} a `clump') moving inwards through the disc on the viscous timescale, in the case of \subref{fig:disc_corona:point} a compact, point-like corona, close to the black hole, where the accreting clump modulates the seed photon flux incident upon the corona before directly influencing the structure of the corona when it reaches the inner disc, and \subref{fig:disc_corona:ext} the two-component corona, proposed by \citet{1zw1_corona_paper}, where the accreting clump modules first the X-ray flux from the low-height component of the corona over the inner disc, before reaching the collimated core of the corona.}
\label{fig:disc_corona}
\end{figure*}

An anti-correlation is also observed, at negative time lags in the correlation function, corresponding to a decrease in reverberation time and scale height of the corona around 20\ks\ before the peak count rate. A similar decrease in scale height is observed before the bright X-ray flares began in 1\,Zw\,1 (Fig.~\ref{fig:1zw1_flare_lag}). There are two possible interpretations of this anti-correlation, illustrated in Fig.~\ref{fig:disc_corona}. In the case of a compact, central corona (\textit{i.e.} the lamppost), such an anti-correlation would be produced by a decrease in scale height as the count rate first increases. If the accreting clump produces an increased seed photon flux before it reaches the inner disc and can impart energy to the corona, the enhancement in seed photons will lead to Compton cooling of the corona, reducing its energy, and, in principle, reducing its volume. The ratio of the X-ray luminosity to the internal energy of the corona (estimated from its temperature) is such that the cooling time is short and the corona requires constant energy injection to maintain the X-ray emission \citep{fabian+2015}.

Alternatively, the corona could comprise two components: a centrally-collimated component (\textit{i.e.} the lamppost, akin to the base of a jet) and an extended component over the surface of the inner accretion disc, as proposed by \citet{1zw1_corona_paper}. In this case, the increased X-ray flux comes from the extended disc corona as the clump propagates inwards. Since this component of the X-ray emission is close to the surface of the disc, the \textit{average} scale-height of the corona decreases, while there is no change to the core of the corona. Then, when the clump reaches the inner disc, the scale height of the collimated core of the corona increases (and in the most extreme cases, the core is accelerated upwards during an X-ray flare).

In general relativistic magneto-hydrodynamic (GRMHD) simulations of accretion onto black holes, high-temperature optically thin layers of plasma are seen on the surface of the inner disc, which may be interpreted as the disc component of the corona \citep{liska+2022,liska+2023}, Reconnection of plasmoids between field line anchored to the spinning black hole and to the disc \citep{sridhar+2021} or the collapse of a jet structure in the magnetic field lines anchored to the black hole \citep{yuan_fluxtubes_2} can lead to dissipation of the energy extracted from the spin of the black hole in a compact, central corona, corresponding to the collimated core inferred from spectral timing measurements.

\section{Conclusions}
Spectral timing analyses based on wavelet transforms provide a new means to measure the time variability of the X-ray emission from accreting systems, including AGN, stellar mass black holes, and neutron stars. In particular, wavelet spectral timing analysis can be used to trace the variability of X-ray reverberation signals from the inner accretion disc on short timescales.

Using wavelet spectral timing analysis, the reverberation of the X-ray continuum in the iron K line from the inner accretion disc was detected in the AGN IRAS\,13224$-$3809 and MCG--6-30-15. The iron K reverberation signal had previously been missing in these two AGN, despite their spectra displaying strong, relativistically broadened iron K lines from the inner disc, and reverberation in their soft X-ray emission. The iron K reverberation time is found to scale with the black hole mass in agreement with the remainder of the Seyfert-type AGN sample in which reverberation has been observed and corresponds to the light travel time across a distance of around 5\rg, suggesting a compact corona close to the black hole and inner accretion disc.

The iron K reverberation signal in these two AGN was found to be transitory in nature, and analysis of the X-ray flux spectrum and covariance spectrum reveals that there are time periods in which the iron K reverberation signal is hidden as the variability in the iron K band becomes dominated by changes in ultrafast outflows (UFOs), launched from the inner disc. Reverberation in the iron K band was detectable during approximately 49 per cent of the time in IRAS\,13224$-$3809 and 23 per cent of the time in MCG--6-30-15, with the UFOs dominating the variability the remiander of the time. Wavelet analysis enables signals to be separated from time periods during which either reverberation or the UFO dominates the variability.

Wavelet analysis follows the evolution of the reverberation time between the corona and inner accretion disc as a function of time. During bright X-ray flares launched around the supermassive black hole in the AGN I\,Zw\,1 in 2020, we observe how the reverberation lag time and scale height of the corona increases as the corona is accelerated away from the black hole and accretion disc.

The behaviour giving rise to the continuous, stochastic variability of the corona is more complex. The count rate is correlated with the reverberation timescale and the scale height of the corona, with the corona expanding as the count rate increases. Hysteresis is observed between the count rate and size of the corona, and a time of $10\sim 40$\ks\ is observed between the initial increase in X-ray luminosity and the increase in scale height of the corona. This time lag can be related to the viscous propagation time of variations in the mass accretion rate through the inner 7$\sim$10\rg\ of the disc, and by measuring the relationship between the structure of the corona and the seed photon flux, we can acquire a deeper understanding of the disc-corona connection around accreting black holes.



\section*{Data Availability}
The data used in this study are available in the \textit{XMM-Newton} public archive. \textit{XMM-Newton} observations can be accessed via the XMM Science Archive (\url{http://nxsa.esac.esa.int/nxsa-web}). Analysis of the X-ray spectra was conducted using \textsc{xspec}, distributed as part of the \textsc{heasoft} package (\url{https://heasarc.gsfc.nasa.gov/docs/software/heasoft}). Codes used to perform the analysis are available upon request to the author.

\section*{Acknowledgments}
This work was supported by the NASA Astrophysics Data Analysis Program under grant 80NSSC22K0406. Computing for this project was performed on the Sherlock cluster. DRW thanks Stanford University and the Stanford Research Computing Center for providing computational resources and support.

\appendix
\section{Wavelet lag errors and uncertainties}
\label{app:errors}
To test the analytic approximation of the phase and lag error (Equation~\ref{equ:phase_error}), Monte Carlo simulations were performed to estimate the distribution of lag measurements between the 0.3-1\keV\ and 1.2-4\keV\ bands in IRAS\,13224$-$3809. Each of the observed light curves was resampled 1,000 times by replacing the count rate measured in each bin with a random count rate drawn from a Poisson distribution with mean corresponding to the observed count rate in that bin. The wavelet analyses were then performed on each set of resampled light curves, calculating the lag or coherence spectra for each sample. The distributions of sampled spectra were then used to estimate the confidence limits, and hence the error bars, on each measurement.

\begin{figure*}
\centering
\subfigure[Lag measurements for small time bins] {
\includegraphics[width=0.49\textwidth]{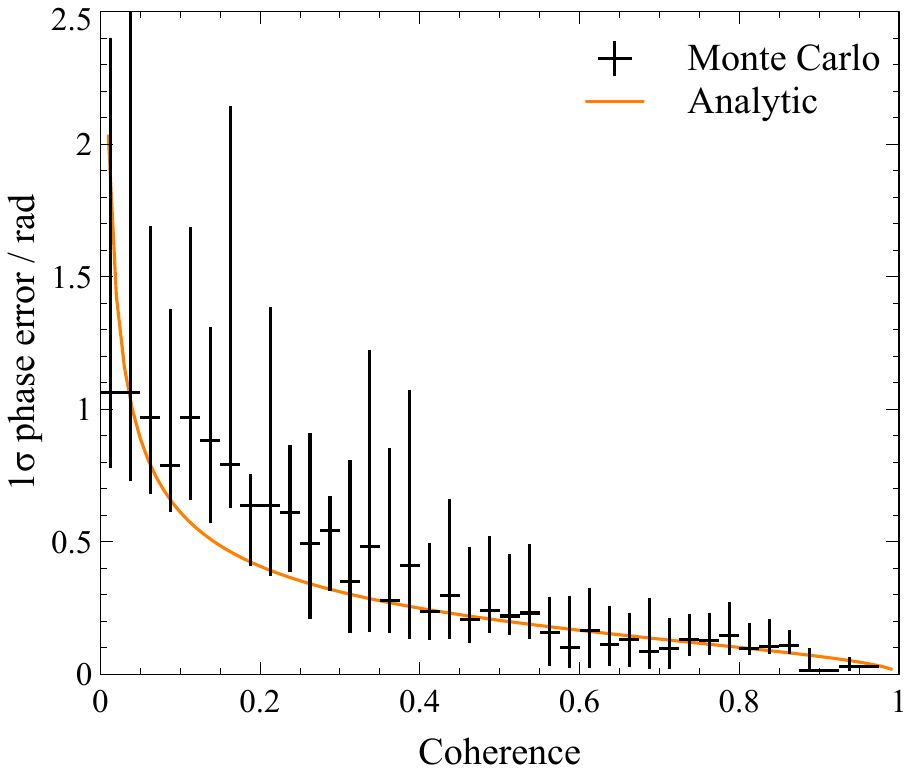}
\label{fig:lag_error:short}
}
\subfigure[Lag measurements averaged over many time bins] {
\includegraphics[width=0.49\textwidth]{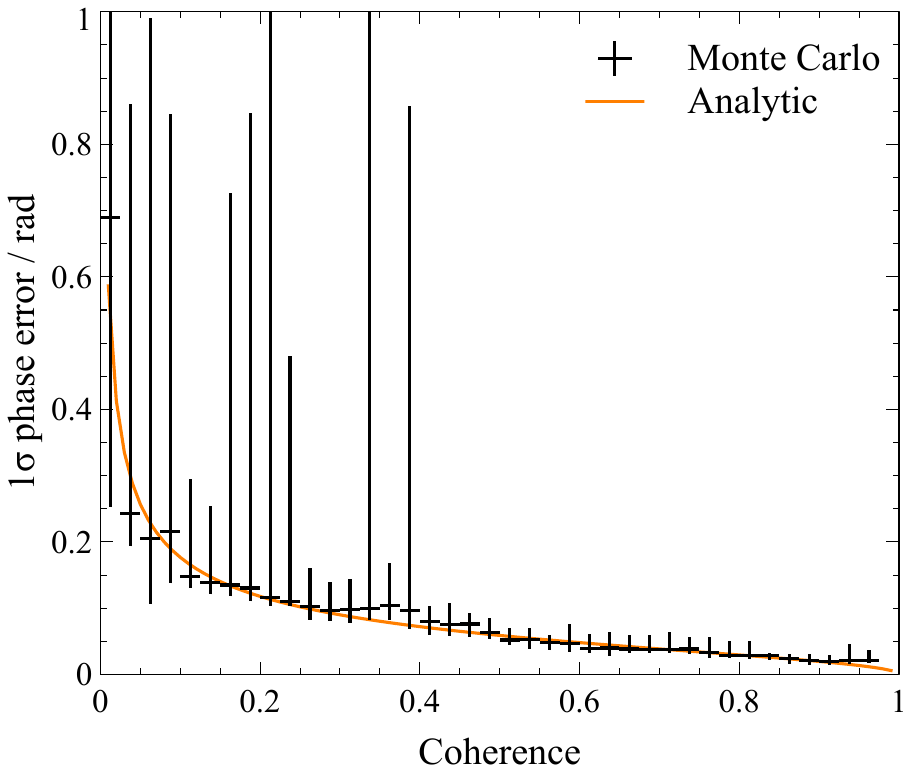}
\label{fig:lag_error:long}
}
\caption[]{Comparison of the uncertainty or error bar on phase lag measurements, $\Delta\varphi$, made from the wavelet coherence function, between Monte Carlo simulations based upon the observed light curves from IRAS\,13224$-$3809 and analytic approximation, for \subref{fig:lag_error:short} the case where the lag is averaged over a small number of time bins, relative to the width of the moving average filter along the time axis, and \subref{fig:lag_error:long} for the case where the lag is averaged over a number of time bins much larger than the width of the moving average filter along the time axis, such as when the lag \textit{vs.} energy spectrum is calculated. Coherence and phase lag uncertainty measurements from 2,126 time bins are binned according to their coherence values. Data points show the median phase error in each bin from the Monte Carlo sample, and the error bars show the $1\sigma$ deviation. The orange line shows the expectation from Equation~\ref{equ:phase_error}, with $N=12$ in \subref{fig:lag_error:short} and $N=12^2=144$ in \subref{fig:lag_error:long}.}
\label{fig:lag_error}
\end{figure*}

Two cases are considered:
\begin{enumerate}
\item When the lag measurement is made in individual time bins, or averaged over a small number of bins relative to the width of the moving average filter used in the wavelet coherence calculation along the time axis (Fig.~\ref{fig:lag_error}a). In this case, the phase lag is approximated using Equation~\ref{equ:phase_error}, with $N$ corresponding to the number of frequency points over which the moving average filter operates ($N=12$).
\item When the lag measurement is averaged over a large number of bins relative to the width of the moving average filter used in the wavelet coherence calculation along the time axis, such as when the lag \textit{vs.} energy spectrum is calculated (Fig.~\ref{fig:lag_error}b). In this case, the phase lag is approximated using Equation~\ref{equ:phase_error}, with $N$ corresponding to the product of the numbers of frequency and time points over which the moving average filter operates ($N=12^2=144$).
\end{enumerate}

Equation~\ref{equ:phase_error} with the appropriate value of $N$ was found to provide a good approximation to the uncertainty on the phase and time lag measurements in the limit of high coherence. When $\gamma^2 > 0.4$, the approximation is good. When $0.25 < \gamma^2 < 0.4$, the approximation to the median phase uncertainty in the sample remains good, however there starts to be significant scatter between the observations. When the coherence is low, the analytic approximation breaks down and cannot be relied upon to estimate the phase uncertainty. When $gamma^2 < 0.25$, the analytic approximation can underestimate the median lag by up to 50 per cent when wavelet analysis is used to calculate lags in small time bins.

\section{Reverberation times for point sources}
\label{app:raytrace}
In order to calculate the characteristic reverberation time lag, fully accounting for relativistic time delays around the black hole and all possible light paths between the corona, the disc and the observer, it is necessary to compute the average arrival time through the impulse response function, weighted by the count rate in each time bin \citep[\textit{e.g.}][]{lag_spectra_paper}.

\begin{figure*}
\centering
\subfigure[] {
\includegraphics[width=0.49\textwidth]{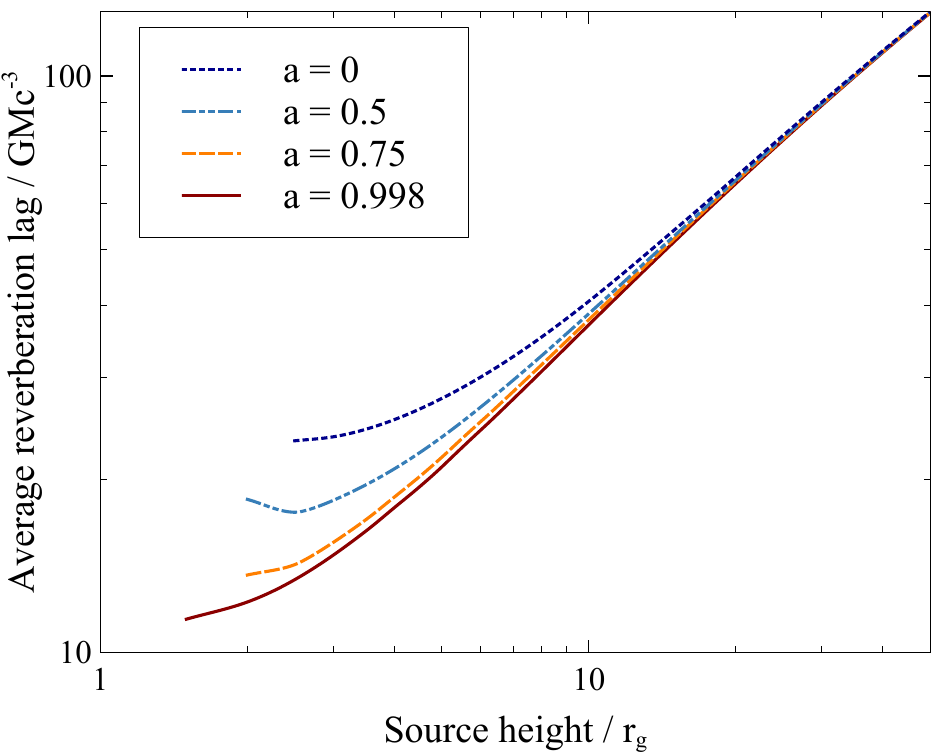}
\label{fig:point_lag:spin}
}
\subfigure[] {
\includegraphics[width=0.49\textwidth]{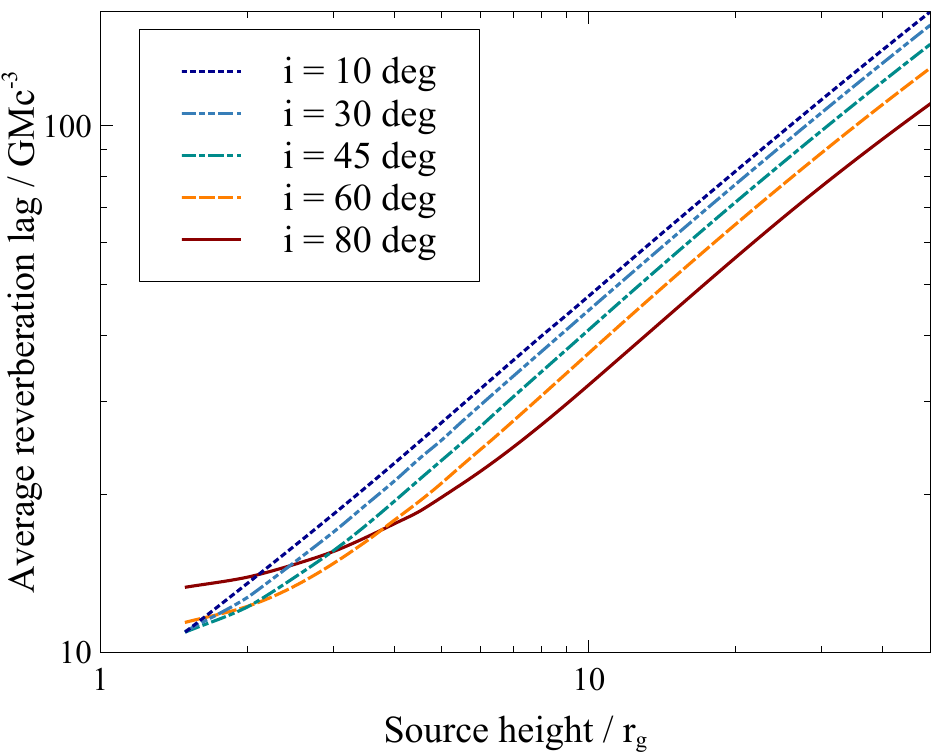}
\label{fig:point_lag:incl}
}
\caption[]{The average reverberation time lag for a point-like corona as a function of coronal height above the black hole, for \subref{fig:point_lag:spin} varying black hole spin parameter, $a = J/Mc$, in units of $GMc^{-2}$, and \subref{fig:point_lag:incl} accretion discs viewed at different inclinations, with 0\,deg corresponding to a disc viewed edge-on, each for the case of a maximally-spinning black hole with $a=0.998$.. The average lag is calculated from the weighted average of the impulse response functions, derived from general relativistic ray tracing simulations of X-ray reverberation around black holes.}
\label{fig:point_lag}
\end{figure*}

Fig.~\ref{fig:point_lag} shows the average lag time for the simplest case of a point-like corona above as a function of the height of the corona above the black hole, for different values of the different black hole spin, and different inclination angles at which the accretion disc is viewed. The impulse response functions were calculated using the \textsc{CUDAKerr} general relativistic ray tracing code \citep{understanding_emis_paper,propagating_lag_paper}.

These plots may be used as a reference to estimate the scale height of the corona from the measured reverberation time lag. The plotted values correspond to the `intrinsic' time lag, and the measured lag time will be diluted according to the relative contribution of the continuum and reverberating emission in the energy bands for which the lag is measured (Equation~\ref{equ:dilution}).

For decreasing black hole spin, the reverberation time lags become longer for coron\ae\ at lower heights above the black hole, due to the increasing radius of the innermost stable orbit. For decreasing spin, a low-height corona is further from the inner edge of the disc. For larger coronal heights, though, the reverberation signal is dominated by the outer disc, thus the lag becomes independent of the spin.

The variation in reverberation time lag as a function of inclination as a geometric effect. For a disc viewed edge-on, the light path from the corona to the disc to the observer is minimised when reverberation signal comes from relatively large radius (since the ray travelling between the corona and disc is coming towards the observer). When the disc is viewed face-on, the total light path is longer since the ray is travelling away from the observer as it propagates from the corona to the disc. This effect is negated for the smallest coronal heights, however, by the gravitational time delays experienced by the rays travelling to the inner disc (which dominate the reverberation signal for point sources at low height).

\bibliographystyle{mnras}
\bibliography{agn}

\begin{thebibliography}{}
\makeatletter
\relax
\def\mn@urlcharsother{\let\do\@makeother \do\$\do\&\do\#\do\^\do\_\do\%\do\~}
\def\mn@doi{\begingroup\mn@urlcharsother \@ifnextchar [ {\mn@doi@}
  {\mn@doi@[]}}
\def\mn@doi@[#1]#2{\def\@tempa{#1}\ifx\@tempa\@empty \href
  {http://dx.doi.org/#2} {doi:#2}\else \href {http://dx.doi.org/#2} {#1}\fi
  \endgroup}
\def\mn@eprint#1#2{\mn@eprint@#1:#2::\@nil}
\def\mn@eprint@arXiv#1{\href {http://arxiv.org/abs/#1} {{\tt arXiv:#1}}}
\def\mn@eprint@dblp#1{\href {http://dblp.uni-trier.de/rec/bibtex/#1.xml}
  {dblp:#1}}
\def\mn@eprint@#1:#2:#3:#4\@nil{\def\@tempa {#1}\def\@tempb {#2}\def\@tempc
  {#3}\ifx \@tempc \@empty \let \@tempc \@tempb \let \@tempb \@tempa \fi \ifx
  \@tempb \@empty \def\@tempb {arXiv}\fi \@ifundefined
  {mn@eprint@\@tempb}{\@tempb:\@tempc}{\expandafter \expandafter \csname
  mn@eprint@\@tempb\endcsname \expandafter{\@tempc}}}

\bibitem[\protect\citeauthoryear{{Alston} et~al.,}{{Alston}
  et~al.}{2019}]{alston_iras1}
{Alston} W.~N.,  et~al., 2019, \mn@doi [\mnras] {10.1093/mnras/sty2527}, \href
  {https://ui.adsabs.harvard.edu/abs/2019MNRAS.482.2088A} {482, 2088}

\bibitem[\protect\citeauthoryear{{Alston} et~al.,}{{Alston}
  et~al.}{2020}]{alston_iras2}
{Alston} W.~N.,  et~al., 2020, \mn@doi [Nature Astronomy]
  {10.1038/s41550-019-1002-x}, \href
  {https://ui.adsabs.harvard.edu/abs/2020NatAs...4..597A} {4, 597}

\bibitem[\protect\citeauthoryear{{Ar{\'e}valo} \& {Uttley}}{{Ar{\'e}valo} \&
  {Uttley}}{2006}]{arevalo+2006}
{Ar{\'e}valo} P.,  {Uttley} P.,  2006, \mn@doi [\mnras]
  {10.1111/j.1365-2966.2006.09989.x}, \href
  {http://adsabs.harvard.edu/abs/2006MNRAS.367..801A} {367, 801}

\bibitem[\protect\citeauthoryear{Bendat \& Piersol}{Bendat \&
  Piersol}{2011}]{bendat_piersol}
Bendat J.~S.,  Piersol A.~G.,  2011, {Random data: analysis and measurement
  procedures}.
John Wiley \& Sons

\bibitem[\protect\citeauthoryear{{Blandford} \& {Payne}}{{Blandford} \&
  {Payne}}{1982}]{blandford_payne}
{Blandford} R.~D.,  {Payne} D.~G.,  1982, \mnras, \href
  {http://adsabs.harvard.edu/abs/1982MNRAS.199..883B} {199, 883}

\bibitem[\protect\citeauthoryear{{Brenneman} \& {Reynolds}}{{Brenneman} \&
  {Reynolds}}{2006}]{brenneman_reynolds}
{Brenneman} L.~W.,  {Reynolds} C.~S.,  2006, \mn@doi [\apj] {10.1086/508146},
  \href {http://adsabs.harvard.edu/abs/2006ApJ...652.1028B} {652, 1028}

\bibitem[\protect\citeauthoryear{{Caballero-Garc{\'\i}a}, {Papadakis}, {Dov{\v
  c}iak}, {Bursa}, {Epitropakis}, {Karas}  \&
  {Svoboda}}{{Caballero-Garc{\'\i}a} et~al.}{2018}]{caballero-garcia+2018}
{Caballero-Garc{\'\i}a} M.~D.,  {Papadakis} I.~E.,  {Dov{\v c}iak} M.,  {Bursa}
  M.,  {Epitropakis} A.,  {Karas} V.,   {Svoboda} J.,  2018, \mn@doi [\mnras]
  {10.1093/mnras/sty1990}, \href
  {https://ui.adsabs.harvard.edu/abs/2018MNRAS.480.2650C} {480, 2650}

\bibitem[\protect\citeauthoryear{{Caballero-Garc{\'\i}a}, {Papadakis}, {Dov{\v
  c}iak}, {Bursa}, {Svoboda}  \& {Karas}}{{Caballero-Garc{\'\i}a}
  et~al.}{2020}]{caballero-garcia+2020}
{Caballero-Garc{\'\i}a} M.~D.,  {Papadakis} I.~E.,  {Dov{\v c}iak} M.,  {Bursa}
  M.,  {Svoboda} J.,   {Karas} V.,  2020, \mn@doi [\mnras]
  {10.1093/mnras/staa2554}, \href
  {https://ui.adsabs.harvard.edu/abs/2020MNRAS.498.3184C} {498, 3184}

\bibitem[\protect\citeauthoryear{{Cackett}, {Zoghbi}, {Reynolds}, {Fabian},
  {Kara}, {Uttley}  \& {Wilkins}}{{Cackett} et~al.}{2014}]{cackett_ngc4151}
{Cackett} E.~M.,  {Zoghbi} A.,  {Reynolds} C.,  {Fabian} A.~C.,  {Kara} E.,
  {Uttley} P.,   {Wilkins} D.~R.,  2014, \mn@doi [\mnras]
  {10.1093/mnras/stt2424}, \href
  {http://adsabs.harvard.edu/abs/2014MNRAS.438.2980C} {438, 2980}

\bibitem[\protect\citeauthoryear{{Cackett}, {Bentz}  \& {Kara}}{{Cackett}
  et~al.}{2021}]{cackett_reverb_review}
{Cackett} E.~M.,  {Bentz} M.~C.,   {Kara} E.,  2021, \mn@doi [iScience]
  {10.1016/j.isci.2021.102557}, \href
  {https://ui.adsabs.harvard.edu/abs/2021iSci...24j2557C} {24, 102557}

\bibitem[\protect\citeauthoryear{{Czerny}, {Lachowicz}, {Dov{\v c}iak},
  {Karas}, {Pech{\'a}{\v c}ek}  \& {Das}}{{Czerny} et~al.}{2010}]{czerny+2010}
{Czerny} B.,  {Lachowicz} P.,  {Dov{\v c}iak} M.,  {Karas} V.,  {Pech{\'a}{\v
  c}ek} T.,   {Das} T.~K.,  2010, \mn@doi [\aap] {10.1051/0004-6361/200913724},
  \href {https://ui.adsabs.harvard.edu/abs/2010A\&A...524A..26C} {524, A26}

\bibitem[\protect\citeauthoryear{{De Marco}, {Ponti}, {Cappi}, {Dadina},
  {Uttley}, {Cackett}, {Fabian}  \& {Miniutti}}{{De Marco}
  et~al.}{2013}]{demarco+2012}
{De Marco} B.,  {Ponti} G.,  {Cappi} M.,  {Dadina} M.,  {Uttley} P.,  {Cackett}
  E.~M.,  {Fabian} A.~C.,   {Miniutti} G.,  2013, \mn@doi [\mnras]
  {10.1093/mnras/stt339}, \href
  {http://adsabs.harvard.edu/abs/2013MNRAS.431.2441D} {431, 2441}

\bibitem[\protect\citeauthoryear{{De Marco}, {Ponti}, {Mu{\~n}oz-Darias}  \&
  {Nandra}}{{De Marco} et~al.}{2015}]{demarco+2015}
{De Marco} B.,  {Ponti} G.,  {Mu{\~n}oz-Darias} T.,   {Nandra} K.,  2015,
  \mn@doi [\apj] {10.1088/0004-637X/814/1/50}, \href
  {https://ui.adsabs.harvard.edu/abs/2015ApJ...814...50D} {814, 50}

\bibitem[\protect\citeauthoryear{{Done}, {Davis}, {Jin}, {Blaes}  \&
  {Ward}}{{Done} et~al.}{2012}]{done_jin}
{Done} C.,  {Davis} S.~W.,  {Jin} C.,  {Blaes} O.,   {Ward} M.,  2012, \mn@doi
  [\mnras] {10.1111/j.1365-2966.2011.19779.x}, \href
  {https://ui.adsabs.harvard.edu/abs/2012MNRAS.420.1848D} {420, 1848}

\bibitem[\protect\citeauthoryear{{Edelson}, {Turner}, {Pounds}, {Vaughan},
  {Markowitz}, {Marshall}, {Dobbie}  \& {Warwick}}{{Edelson}
  et~al.}{2002}]{edelson_fvar}
{Edelson} R.,  {Turner} T.~J.,  {Pounds} K.,  {Vaughan} S.,  {Markowitz} A.,
  {Marshall} H.,  {Dobbie} P.,   {Warwick} R.,  2002, \mn@doi [\apj]
  {10.1086/323779}, \href {http://adsabs.harvard.edu/abs/2002ApJ...568..610E}
  {568, 610}

\bibitem[\protect\citeauthoryear{{Fabian}, {Rees}, {Stella}  \&
  {White}}{{Fabian} et~al.}{1989}]{fabian+89}
{Fabian} A.~C.,  {Rees} M.~J.,  {Stella} L.,   {White} N.~E.,  1989, \mnras,
  \href {http://ukads.nottingham.ac.uk/abs/1989MNRAS.238..729F} {238, 729}

\bibitem[\protect\citeauthoryear{{Fabian} et~al.,}{{Fabian}
  et~al.}{2009}]{fabian+09}
{Fabian} A.~C.,  et~al., 2009, \mn@doi [\nat] {10.1038/nature08007}, \href
  {http://adsabs.harvard.edu/abs/2009Natur.459..540F} {459, 540}

\bibitem[\protect\citeauthoryear{{Fabian} et~al.,}{{Fabian}
  et~al.}{2012}]{1h0707_jan11}
{Fabian} A.~C.,  et~al., 2012, \mn@doi [\mnras]
  {10.1111/j.1365-2966.2011.19676.x}, \href
  {http://adsabs.harvard.edu/abs/2012MNRAS.419..116F} {419, 116}

\bibitem[\protect\citeauthoryear{{Fabian} et~al.,}{{Fabian}
  et~al.}{2013}]{fabian+2013}
{Fabian} A.~C.,  et~al., 2013, \mn@doi [\mnras] {10.1093/mnras/sts504}, \href
  {https://ui.adsabs.harvard.edu/abs/2013MNRAS.429.2917F} {429, 2917}

\bibitem[\protect\citeauthoryear{{Fabian}, {Lohfink}, {Kara}, {Parker},
  {Vasudevan}  \& {Reynolds}}{{Fabian} et~al.}{2015}]{fabian+2015}
{Fabian} A.~C.,  {Lohfink} A.,  {Kara} E.,  {Parker} M.~L.,  {Vasudevan} R.,
  {Reynolds} C.~S.,  2015, \mn@doi [\mnras] {10.1093/mnras/stv1218}, \href
  {https://ui.adsabs.harvard.edu/abs/2015MNRAS.451.4375F} {451, 4375}

\bibitem[\protect\citeauthoryear{{Freeman}, {Kashyap}, {Rosner}  \&
  {Lamb}}{{Freeman} et~al.}{2002}]{wavdetect}
{Freeman} P.~E.,  {Kashyap} V.,  {Rosner} R.,   {Lamb} D.~Q.,  2002, \mn@doi
  [\apjs] {10.1086/324017}, \href
  {https://ui.adsabs.harvard.edu/abs/2002ApJS..138..185F} {138, 185}

\bibitem[\protect\citeauthoryear{{Galeev}, {Rosner}  \& {Vaiana}}{{Galeev}
  et~al.}{1979}]{galeev+79}
{Galeev} A.~A.,  {Rosner} R.,   {Vaiana} G.~S.,  1979, \mn@doi [\apj]
  {10.1086/156957}, \href {http://adsabs.harvard.edu/abs/1979ApJ...229..318G}
  {229, 318}

\bibitem[\protect\citeauthoryear{{George} \& {Fabian}}{{George} \&
  {Fabian}}{1991}]{george_fabian}
{George} I.~M.,  {Fabian} A.~C.,  1991, \mnras, \href
  {http://ukads.nottingham.ac.uk/abs/1991MNRAS.249..352G} {249, 352}

\bibitem[\protect\citeauthoryear{{Ghosh}, {Gallo}  \& {Gonzalez}}{{Ghosh}
  et~al.}{2023}]{ghosh+2023}
{Ghosh} A.,  {Gallo} L.~C.,   {Gonzalez} A.~G.,  2023, \mn@doi [\mnras]
  {10.1093/mnras/stad1955}, \href
  {https://ui.adsabs.harvard.edu/abs/2023MNRAS.524.1478G} {524, 1478}

\bibitem[\protect\citeauthoryear{{Gonz{\'a}lez-Mart{\'\i}n} \&
  {Vaughan}}{{Gonz{\'a}lez-Mart{\'\i}n} \&
  {Vaughan}}{2012}]{gonzalezmartin+2012}
{Gonz{\'a}lez-Mart{\'\i}n} O.,  {Vaughan} S.,  2012, \mn@doi [\aap]
  {10.1051/0004-6361/201219008}, \href
  {https://ui.adsabs.harvard.edu/abs/2012A\&A...544A..80G} {544, A80}

\bibitem[\protect\citeauthoryear{{Haardt} \& {Maraschi}}{{Haardt} \&
  {Maraschi}}{1991}]{haardt+91}
{Haardt} F.,  {Maraschi} L.,  1991, \mn@doi [\apjl] {10.1086/186171}, \href
  {http://adsabs.harvard.edu/abs/1991ApJ...380L..51H} {380, L51}

\bibitem[\protect\citeauthoryear{{Igo} et~al.,}{{Igo} et~al.}{2020}]{igo+2020}
{Igo} Z.,  et~al., 2020, \mn@doi [\mnras] {10.1093/mnras/staa265}, \href
  {https://ui.adsabs.harvard.edu/abs/2020MNRAS.493.1088I} {493, 1088}

\bibitem[\protect\citeauthoryear{{Jiang} et~al.,}{{Jiang}
  et~al.}{2018}]{jiang_iras}
{Jiang} J.,  et~al., 2018, \mn@doi [\mnras] {10.1093/mnras/sty836}, \href
  {https://ui.adsabs.harvard.edu/\#abs/2018MNRAS.477.3711J} {477, 3711}

\bibitem[\protect\citeauthoryear{{Kammoun} \& {Papadakis}}{{Kammoun} \&
  {Papadakis}}{2017}]{kammoun+2017}
{Kammoun} E.~S.,  {Papadakis} I.~E.,  2017, \mn@doi [\mnras]
  {10.1093/mnras/stx2181}, \href
  {https://ui.adsabs.harvard.edu/abs/2017MNRAS.472.3131K} {472, 3131}

\bibitem[\protect\citeauthoryear{{Kara}, {Fabian}, {Cackett}, {Uttley},
  {Wilkins}  \& {Zoghbi}}{{Kara} et~al.}{2013}]{kara+13}
{Kara} E.,  {Fabian} A.~C.,  {Cackett} E.~M.,  {Uttley} P.,  {Wilkins} D.~R.,
  {Zoghbi} A.,  2013, \mn@doi [\mnras] {10.1093/mnras/stt1055}, \href
  {http://adsabs.harvard.edu/abs/2013MNRAS.434.1129K} {434, 1129}

\bibitem[\protect\citeauthoryear{{Kara} et~al.,}{{Kara}
  et~al.}{2014}]{kara+2014}
{Kara} E.,  et~al., 2014, \mn@doi [\mnras] {10.1093/mnras/stu1750}, \href
  {https://ui.adsabs.harvard.edu/abs/2014MNRAS.445...56K} {445, 56}

\bibitem[\protect\citeauthoryear{{Kara}, {Alston}, {Fabian}, {Cackett},
  {Uttley}, {Reynolds}  \& {Zoghbi}}{{Kara} et~al.}{2016}]{kara_global}
{Kara} E.,  {Alston} W.~N.,  {Fabian} A.~C.,  {Cackett} E.~M.,  {Uttley} P.,
  {Reynolds} C.~S.,   {Zoghbi} A.,  2016, \mn@doi [\mnras]
  {10.1093/mnras/stw1695}, \href
  {https://ui.adsabs.harvard.edu/\#abs/2016MNRAS.462..511K} {462, 511}

\bibitem[\protect\citeauthoryear{{Kara} et~al.,}{{Kara}
  et~al.}{2019}]{kara+2019}
{Kara} E.,  et~al., 2019, \mn@doi [\nat] {10.1038/s41586-018-0803-x}, \href
  {https://ui.adsabs.harvard.edu/abs/2019Natur.565..198K} {565, 198}

\bibitem[\protect\citeauthoryear{{Kotov}, {Churazov}  \& {Gilfanov}}{{Kotov}
  et~al.}{2001}]{kotov+2001}
{Kotov} O.,  {Churazov} E.,   {Gilfanov} M.,  2001, \mn@doi [\mnras]
  {10.1046/j.1365-8711.2001.04769.x}, \href
  {http://adsabs.harvard.edu/abs/2001MNRAS.327..799K} {327, 799}

\bibitem[\protect\citeauthoryear{{Lachowicz} \& {Czerny}}{{Lachowicz} \&
  {Czerny}}{2005}]{lachowicz+2005}
{Lachowicz} P.,  {Czerny} B.,  2005, \mn@doi [\mnras]
  {10.1111/j.1365-2966.2005.09197.x}, \href
  {https://ui.adsabs.harvard.edu/abs/2005MNRAS.361..645L} {361, 645}

\bibitem[\protect\citeauthoryear{Lau \& Weng}{Lau \& Weng}{1995}]{lau+1995}
Lau K.-M.,  Weng H.,  1995, \mn@doi [Bulletin of the American Meteorological
  Society] {10.1175/1520-0477(1995)076<2391:CSDUWT>2.0.CO;2}, 76, 2391

\bibitem[\protect\citeauthoryear{{Liska}, {Musoke}, {Tchekhovskoy}, {Porth}  \&
  {Beloborodov}}{{Liska} et~al.}{2022}]{liska+2022}
{Liska} M.~T.~P.,  {Musoke} G.,  {Tchekhovskoy} A.,  {Porth} O.,
  {Beloborodov} A.~M.,  2022, \mn@doi [\apjl] {10.3847/2041-8213/ac84db}, \href
  {https://ui.adsabs.harvard.edu/abs/2022ApJ...935L...1L} {935, L1}

\bibitem[\protect\citeauthoryear{{Liska}, {Hesp}, {Tchekhovskoy}, {Ingram},
  {van der Klis}  \& {Markoff}}{{Liska} et~al.}{2023}]{liska+2023}
{Liska} M.,  {Hesp} C.,  {Tchekhovskoy} A.,  {Ingram} A.,  {van der Klis} M.,
  {Markoff} S.~B.,  2023, \mn@doi [\na] {10.1016/j.newast.2023.102012}, \href
  {https://ui.adsabs.harvard.edu/abs/2023NewA..10102012L} {101, 102012}

\bibitem[\protect\citeauthoryear{{Mallick}, {Wilkins}, {Alston}, {Markowitz},
  {De Marco}, {Parker}, {Lohfink}  \& {Stalin}}{{Mallick}
  et~al.}{2021}]{mallick+2021}
{Mallick} L.,  {Wilkins} D.~R.,  {Alston} W.~N.,  {Markowitz} A.,  {De Marco}
  B.,  {Parker} M.~L.,  {Lohfink} A.~M.,   {Stalin} C.~S.,  2021, \mn@doi
  [\mnras] {10.1093/mnras/stab627}, \href
  {https://ui.adsabs.harvard.edu/abs/2021MNRAS.503.3775M} {503, 3775}

\bibitem[\protect\citeauthoryear{{Marinucci} et~al.,}{{Marinucci}
  et~al.}{2014}]{marinucci+2014}
{Marinucci} A.,  et~al., 2014, \mn@doi [\apj] {10.1088/0004-637X/787/1/83},
  \href {https://ui.adsabs.harvard.edu/abs/2014ApJ...787...83M} {787, 83}

\bibitem[\protect\citeauthoryear{{Miyamoto}, {Kitamoto}, {Mitsuda}  \&
  {Dotani}}{{Miyamoto} et~al.}{1988}]{miyamoto+88}
{Miyamoto} S.,  {Kitamoto} S.,  {Mitsuda} K.,   {Dotani} T.,  1988, \mn@doi
  [\nat] {10.1038/336450a0}, \href
  {http://adsabs.harvard.edu/abs/1988Natur.336..450M} {336, 450}

\bibitem[\protect\citeauthoryear{{Nowak}, {Vaughan}, {Wilms}, {Dove}  \&
  {Begelman}}{{Nowak} et~al.}{1999}]{nowak+99}
{Nowak} M.~A.,  {Vaughan} B.~A.,  {Wilms} J.,  {Dove} J.~B.,   {Begelman}
  M.~C.,  1999, \mn@doi [\apj] {10.1086/306610}, \href
  {http://adsabs.harvard.edu/abs/1999ApJ...510..874N} {510, 874}

\bibitem[\protect\citeauthoryear{Olhede \& Walden}{Olhede \&
  Walden}{2002}]{olhede+2002}
Olhede S.,  Walden A.,  2002, \mn@doi [IEEE Transactions on Signal Processing]
  {10.1109/TSP.2002.804066}, 50, 2661

\bibitem[\protect\citeauthoryear{{Parker}, {Wilkins}, {Fabian}, {Grupe},
  {Dauser}, {Matt}  \& {Harrison}}{{Parker} et~al.}{2014}]{parker_mrk335}
{Parker} M.~L.,  {Wilkins} D.~R.,  {Fabian} A.~C.,  {Grupe} D.,  {Dauser} T.,
  {Matt} G.,   {Harrison} F.~A.,  2014, \mn@doi [\mnras]
  {10.1093/mnras/stu1246}, \href
  {http://adsabs.harvard.edu/abs/2014MNRAS.443.1723P} {443, 1723}

\bibitem[\protect\citeauthoryear{{Parker} et~al.,}{{Parker}
  et~al.}{2017}]{parker_iras_nature}
{Parker} M.~L.,  et~al., 2017, \mn@doi [\nat] {10.1038/nature21385}, \href
  {https://ui.adsabs.harvard.edu/abs/2017Natur.543...83P} {543, 83}

\bibitem[\protect\citeauthoryear{{Reynolds}}{{Reynolds}}{2021}]{reynolds_spin_review}
{Reynolds} C.~S.,  2021, \mn@doi [\araa] {10.1146/annurev-astro-112420-035022},
  \href {https://ui.adsabs.harvard.edu/abs/2021ARA\&A..59..117R} {59, 117}

\bibitem[\protect\citeauthoryear{Shapiro}{Shapiro}{1964}]{shapiro}
Shapiro I.~I.,  1964, \mn@doi [Phys. Rev. Lett.] {10.1103/PhysRevLett.13.789},
  13, 789

\bibitem[\protect\citeauthoryear{{Silva}, {Uttley}  \& {Costantini}}{{Silva}
  et~al.}{2016}]{silva+2016}
{Silva} C.~V.,  {Uttley} P.,   {Costantini} E.,  2016, \mn@doi [\aap]
  {10.1051/0004-6361/201628555}, \href
  {https://ui.adsabs.harvard.edu/abs/2016A\&A...596A..79S} {596, A79}

\bibitem[\protect\citeauthoryear{{Sridhar}, {Sironi}  \&
  {Beloborodov}}{{Sridhar} et~al.}{2021}]{sridhar+2021}
{Sridhar} N.,  {Sironi} L.,   {Beloborodov} A.~M.,  2021, \mn@doi [\mnras]
  {10.1093/mnras/stab2534}, \href
  {https://ui.adsabs.harvard.edu/abs/2021MNRAS.507.5625S} {507, 5625}

\bibitem[\protect\citeauthoryear{{Tanaka} et~al.,}{{Tanaka}
  et~al.}{1995}]{tanaka+95}
{Tanaka} Y.,  et~al., 1995, \mn@doi [\nat] {10.1038/375659a0}, \href
  {http://adsabs.harvard.edu/abs/1995Natur.375..659T} {375, 659}

\bibitem[\protect\citeauthoryear{{Taylor} \& {Reynolds}}{{Taylor} \&
  {Reynolds}}{2018}]{taylor_reynolds}
{Taylor} C.,  {Reynolds} C.~S.,  2018, \mn@doi [\apj]
  {10.3847/1538-4357/aaad63}, \href
  {https://ui.adsabs.harvard.edu/\#abs/2018ApJ...855..120T} {855, 120}

\bibitem[\protect\citeauthoryear{{Uttley}, {Cackett}, {Fabian}, {Kara}  \&
  {Wilkins}}{{Uttley} et~al.}{2014}]{reverb_review}
{Uttley} P.,  {Cackett} E.~M.,  {Fabian} A.~C.,  {Kara} E.,   {Wilkins} D.~R.,
  2014, \mn@doi [\aapr] {10.1007/s00159-014-0072-0}, \href
  {https://ui.adsabs.harvard.edu/abs/2014A\&ARv..22...72U} {22, 72}

\bibitem[\protect\citeauthoryear{{Vestergaard} \& {Peterson}}{{Vestergaard} \&
  {Peterson}}{2006}]{vestergaard+06}
{Vestergaard} M.,  {Peterson} B.~M.,  2006, \mn@doi [\apj] {10.1086/500572},
  \href {http://adsabs.harvard.edu/abs/2006ApJ...641..689V} {641, 689}

\bibitem[\protect\citeauthoryear{{Wang} et~al.,}{{Wang}
  et~al.}{2022}]{wang_reverb_machine}
{Wang} J.,  et~al., 2022, \mn@doi [\apj] {10.3847/1538-4357/ac6262}, \href
  {https://ui.adsabs.harvard.edu/abs/2022ApJ...930...18W} {930, 18}

\bibitem[\protect\citeauthoryear{{Wilkins} \& {Fabian}}{{Wilkins} \&
  {Fabian}}{2012}]{understanding_emis_paper}
{Wilkins} D.~R.,  {Fabian} A.~C.,  2012, \mn@doi [\mnras]
  {10.1111/j.1365-2966.2012.21308.x}, \href
  {http://adsabs.harvard.edu/abs/2012MNRAS.424.1284W} {424, 1284}

\bibitem[\protect\citeauthoryear{{Wilkins} \& {Fabian}}{{Wilkins} \&
  {Fabian}}{2013}]{lag_spectra_paper}
{Wilkins} D.~R.,  {Fabian} A.~C.,  2013, \mn@doi [\mnras]
  {10.1093/mnras/sts591}, \href
  {http://adsabs.harvard.edu/abs/2013MNRAS.430..247W} {430, 247}

\bibitem[\protect\citeauthoryear{{Wilkins} \& {Gallo}}{{Wilkins} \&
  {Gallo}}{2015}]{mrk335_corona_paper}
{Wilkins} D.~R.,  {Gallo} L.~C.,  2015, \mn@doi [\mnras]
  {10.1093/mnras/stv162}, 449, 129

\bibitem[\protect\citeauthoryear{{Wilkins}, {Kara}, {Fabian}  \&
  {Gallo}}{{Wilkins} et~al.}{2014}]{1h0707_var_paper}
{Wilkins} D.~R.,  {Kara} E.,  {Fabian} A.~C.,   {Gallo} L.~C.,  2014, \mn@doi
  [\mnras] {10.1093/mnras/stu1273}, \href
  {http://adsabs.harvard.edu/abs/2014MNRAS.443.2746W} {443, 2746}

\bibitem[\protect\citeauthoryear{{Wilkins}, {Gallo}, {Grupe}, {Bonson},
  {Komossa}  \& {Fabian}}{{Wilkins} et~al.}{2015}]{mrk335_flare_paper}
{Wilkins} D.~R.,  {Gallo} L.~C.,  {Grupe} D.,  {Bonson} K.,  {Komossa} S.,
  {Fabian} A.~C.,  2015, \mn@doi [\mnras] {10.1093/mnras/stv2130}, \href
  {http://adsabs.harvard.edu/abs/2015MNRAS.454.4440W} {454, 4440}

\bibitem[\protect\citeauthoryear{{Wilkins}, {Cackett}, {Fabian}  \&
  {Reynolds}}{{Wilkins} et~al.}{2016}]{propagating_lag_paper}
{Wilkins} D.~R.,  {Cackett} E.~M.,  {Fabian} A.~C.,   {Reynolds} C.~S.,  2016,
  \mn@doi [\mnras] {10.1093/mnras/stw276}, \href
  {http://adsabs.harvard.edu/abs/2016MNRAS.458..200W} {458, 200}

\bibitem[\protect\citeauthoryear{{Wilkins}, {Gallo}, {Silva}, {Costantini},
  {Brandt}  \& {Kriss}}{{Wilkins} et~al.}{2017}]{1zw1_corona_paper}
{Wilkins} D.~R.,  {Gallo} L.~C.,  {Silva} C.~V.,  {Costantini} E.,  {Brandt}
  W.~N.,   {Kriss} G.~A.,  2017, \mn@doi [\mnras] {10.1093/mnras/stx1814},
  \href {https://ui.adsabs.harvard.edu/\#abs/2017MNRAS.471.4436W} {471, 4436}

\bibitem[\protect\citeauthoryear{{Wilkins}, {Gallo}, {Costantini}, {Brandt}  \&
  {Blandford}}{{Wilkins} et~al.}{2021}]{1zw1_nature}
{Wilkins} D.~R.,  {Gallo} L.~C.,  {Costantini} E.,  {Brandt} W.~N.,
  {Blandford} R.~D.,  2021, \mn@doi [\nat] {10.1038/s41586-021-03667-0}, \href
  {https://ui.adsabs.harvard.edu/abs/2021Natur.595..657W} {595, 657}

\bibitem[\protect\citeauthoryear{{Wilkins}, {Gallo}, {Costantini}, {Brandt}  \&
  {Blandford}}{{Wilkins} et~al.}{2022}]{1zw1_flare_paper}
{Wilkins} D.~R.,  {Gallo} L.~C.,  {Costantini} E.,  {Brandt} W.~N.,
  {Blandford} R.~D.,  2022, \mn@doi [\mnras] {10.1093/mnras/stac416}, \href
  {https://ui.adsabs.harvard.edu/abs/2022MNRAS.512..761W} {512, 761}

\bibitem[\protect\citeauthoryear{{Yuan}, {Spitkovsky}, {Blandford}  \&
  {Wilkins}}{{Yuan} et~al.}{2019}]{yuan_fluxtubes_2}
{Yuan} Y.,  {Spitkovsky} A.,  {Blandford} R.~D.,   {Wilkins} D.~R.,  2019,
  \mn@doi [\mnras] {10.1093/mnras/stz1599}, \href
  {https://ui.adsabs.harvard.edu/abs/2019MNRAS.487.4114Y} {487, 4114}

\bibitem[\protect\citeauthoryear{{Zoghbi}, {Fabian}, {Uttley}, {Miniutti},
  {Gallo}, {Reynolds}, {Miller}  \& {Ponti}}{{Zoghbi} et~al.}{2010}]{zoghbi+09}
{Zoghbi} A.,  {Fabian} A.~C.,  {Uttley} P.,  {Miniutti} G.,  {Gallo} L.~C.,
  {Reynolds} C.~S.,  {Miller} J.~M.,   {Ponti} G.,  2010, \mn@doi [\mnras]
  {10.1111/j.1365-2966.2009.15816.x}, \href
  {http://ukads.nottingham.ac.uk/abs/2010MNRAS.401.2419Z} {401, 2419}

\bibitem[\protect\citeauthoryear{{Zoghbi}, {Fabian}, {Reynolds}  \&
  {Cackett}}{{Zoghbi} et~al.}{2012}]{zoghbi+2012}
{Zoghbi} A.,  {Fabian} A.~C.,  {Reynolds} C.~S.,   {Cackett} E.~M.,  2012,
  \mn@doi [\mnras] {10.1111/j.1365-2966.2012.20587.x}, \href
  {http://adsabs.harvard.edu/abs/2012MNRAS.422..129Z} {422, 129}

\bibitem[\protect\citeauthoryear{{Zoghbi}, {Reynolds}  \& {Cackett}}{{Zoghbi}
  et~al.}{2013}]{zoghbi_gaps}
{Zoghbi} A.,  {Reynolds} C.,   {Cackett} E.~M.,  2013, \mn@doi [\apj]
  {10.1088/0004-637X/777/1/24}, \href
  {https://ui.adsabs.harvard.edu/abs/2013ApJ...777...24Z} {777, 24}

\bibitem[\protect\citeauthoryear{{Zoghbi} et~al.,}{{Zoghbi}
  et~al.}{2014}]{zoghbi+2014}
{Zoghbi} A.,  et~al., 2014, \mn@doi [\apj] {10.1088/0004-637X/789/1/56}, \href
  {http://adsabs.harvard.edu/abs/2014ApJ...789...56Z} {789, 56}

\bibitem[\protect\citeauthoryear{{van de Wouwer}, {Scheunders}, {van Dyck}, {de
  Bodt}, {Wuyts}  \& {van de Heyning}}{{van de Wouwer}
  et~al.}{1997}]{wavelet_speech}
{van de Wouwer} G.,  {Scheunders} P.,  {van Dyck} D.,  {de Bodt} M.,  {Wuyts}
  F.,   {van de Heyning} P.~H.,  1997, Fractals, 5, 165

\makeatother
\end{thebibliography}


\bsp	
\label{lastpage}
\end{document}